\documentclass[aps,twocolumn,preprintnumbers,prd,superscriptaddress,nofootinbib,10pt,floatfix]{revtex4-2}
\usepackage[utf8]{inputenc}
\usepackage{graphicx,amssymb,amsmath,amsthm,amsfonts,epsfig,epsf}
\usepackage[usenames,dvipsnames,svgnames,table]{xcolor}
\usepackage{epstopdf}
\usepackage[caption=false]{subfig}
\usepackage{soul}
\definecolor{darkred}{rgb}{0.5,0,0}
\usepackage{bm}
\usepackage{dcolumn}
\usepackage{latexsym}
\usepackage{rotating}
\usepackage{longtable}

\usepackage{enumerate}
\usepackage{tensor,multirow}
\usepackage[normalem]{ulem}
\usepackage{ragged2e}
\usepackage{booktabs}

\usepackage[english]{babel}

\usepackage{amsmath}
\usepackage{graphicx}
\usepackage[colorlinks=true, allcolors=blue]{hyperref}
\usepackage{scalerel,amssymb}
\usepackage{cancel}
\usepackage{tcolorbox}
\usepackage{empheq}
\usepackage{listings}
\usepackage{xcolor}

\begin{document}

\title{Constraint-satisfying binary boson star initial data via XCFC}

\author{Gabriele Palloni}

\affiliation{Departamento de Astronomía y Astrofísica, Universitat de València, Avinguda Vicent 
Andrés Estellés 19, 46100, Burjassot (València), Spain}

\author{Nicolas Sanchis-Gual}
\affiliation{Departamento de Astronomía y Astrofísica, Universitat de València, Avinguda Vicent 
Andrés Estellés 19, 46100, Burjassot (València), Spain}

\author{José A. Font}
\affiliation{Departamento de Astronomía y Astrofísica, Universitat de València, Avinguda Vicent 
Andrés Estellés 19, 46100, Burjassot (València), Spain}
\affiliation{Observatori Astron\`{o}mic, Universitat de Val\`{e}ncia,
C/ Catedr\'{a}tico Jos\'{e} Beltr\'{a}n 2, 46980, Paterna (Val\`{e}ncia), Spain}

\author{Samuel Santos-Pérez}
\affiliation{Departamento de Matemáticas, Universitat de Val\`{e}ncia, Avinguda Vicent 
Andrés Estellés 19, Burjassot (Val\`{e}ncia), Spain}

\author{Isabel Cordero-Carrión}
\affiliation{Departamento de Matemáticas, Universitat de Val\`{e}ncia, Avinguda Vicent 
Andrés Estellés 19, Burjassot (Val\`{e}ncia), Spain}

\author{Pablo Cerdá-Durán}
\affiliation{Departamento de Astronomía y Astrofísica, Universitat de València, Avinguda Vicent 
Andrés Estellés 19, 46100, Burjassot (València), Spain}
\affiliation{Observatori Astron\`{o}mic, Universitat de Val\`{e}ncia,
C/ Catedr\'{a}tico Jos\'{e} Beltr\'{a}n 2, 46980, Paterna (Val\`{e}ncia), Spain}

\author{Claudio Lazarte}
\affiliation{Departamento de Astronomía y Astrofísica, Universitat de València, Avinguda Vicent 
Andrés Estellés 19, 46100, Burjassot (València), Spain}

\email{gabriele.palloni@uv.es}


\begin{abstract}
Numerical-relativity simulations with non-trivial matter configurations require initial data that satisfy the Hamiltonian and momentum constraints of the Einstein equations. We construct constraint-satisfying scalar-field initial data using the eXtended Conformally Flat Condition (XCFC) formalism, in which the matter variables are conformally rescaled and an auxiliary vector field is introduced. In doing so, we overcome the issues of local uniqueness and convergence of the solutions that arise in the second-order elliptic equations associated with the constraints.
Using an iterative solver method, we demonstrate the convergence of the XCFC approach to a solution for several scalar-field matter systems. Those include Gaussian-like profiles, topological torus configurations, and equal-mass boson star binaries. In particular, for the latter case, it is common to employ the superposition of two isolated boson star solutions in order to build the initial data. We show that our formalism significantly improves upon a superposition approach by generating genuinely constraint-satisfying initial data for boson star binaries.
\end{abstract}
\maketitle

\section{Introduction}

Numerical relativity (NR) is a fundamental tool to investigate the  non-linear dynamics of self-gravitating compact objects, in particular in the context of the modellisation of compact binary coalescences and associated gravitational wave emission.
State-of-the-art NR codes are typically based on a $3+1$ decomposition of Einstein’s equations, which allows the reformulation of the spacetime dynamics as an initial value problem suitable for numerical integration~\cite{Alcubierre_2008,gourgoulhon200731formalismbasesnumerical,Baumgarte:2010ndz}.

Depending on how the Einstein constraint equations are treated during the time evolution, numerical simulations can be classified into two main categories.
In \emph{free evolution} schemes, the Hamiltonian and momentum constraints are solved only at the initial time and subsequently monitored during the evolution, with their violation providing a diagnostic of the numerical accuracy~\cite{Alcubierre_2008,Baumgarte:2010ndz}. In contrast, \emph{constrained evolution} approaches explicitly enforce the constraints throughout the simulation by solving an elliptic subsystem alongside the evolution equations~\cite{Andersson:2001kw, Bonazzola_2004}. Nevertheless, independently of the specific formulation employed, all NR simulations require the construction of suitable initial data.
Such data must satisfy, at least initially, the Einstein constraint equations (namely, the Hamiltonian and momentum constraints) in order to represent a physically consistent spacetime configuration.

Specifically, for compact binary mergers, obtaining constraint-satisfying initial data is crucial to avoid artificial spurious effects that could spoil the subsequent evolution and to ensure a physically consistent initial configuration.
For standard compact-object binaries, such as binary black holes, binary neutron stars, and even hybrid black hole-neutron star systems, this problem has been extensively investigated, and several well-established formalisms for the construction of initial data have been developed~\cite{York1979Sources,Cook_2000,PhysRevD.67.044022,PhysRevLett.82.1350,pfeiffer2004initialvalueproblemnumerical,Baumgarte:2010ndz}. For example, to highlight a specific technique, spectral methods have been successfully applied to black hole–neutron star binaries~\cite{PhysRevD.76.124038}, and recent improvements in binary black hole initial data construction have enhanced robustness for high spin and mass-ratio configurations~\cite{Ossokine_2015}. For more recent reviews, see~\cite{Tichy_2016,Aurrekoetxea2025}.

In addition to these two classes of standard astrophysical compact objects, a variety of theoretical, exotic compact and black--hole-mimicking objects have been proposed in the literature to
address open problems in fundamental, high-energy, and particle physics, and have even been studied with NR simulations~\cite{bezares2025exotic}. These models are
often motivated by beyond General Relativity theories, String Theory or extensions of the Standard Model~\cite{arvanitaki2010string,arvanitaki2011exploring,freitas2021ultralight} and have also been
considered as viable dark matter candidates~\cite{Cardoso_2019,Barack_2019}. Among such exotic compact objects, boson stars represent one of the simplest
and most extensively studied examples. Originally proposed in~\cite{PhysRev.172.1331}, boson stars are self-gravitating solitonic configurations supported by scalar (or vector) fields. In recent decades, numerous studies have investigated their physical and astrophysical properties~\cite{PhysRev.187.1767, PhysRevD.56.762, PhysRevD.42.384, PhysRevLett.66.1659, sanchis2017numerical,sanchis2019nonlinear,herdeiro2024non}. Depending on their compactness and internal structure, boson stars can exhibit features commonly associated with black holes, such as ergoregions or light rings, while remaining horizonless objects (see~\cite{Franz_E_Schunck_2003,Liebling_2012, Visinelli_2021} and references therein). This  makes boson stars especially relevant in the context of black-hole mimickers and tests of strong-field gravity. In particular, numerical evolutions of boson star binaries, including their vector counterparts known as Proca stars~\cite{BRITO2016291}, have been reported in several studies. Those have explored different  scenarios, including head-on collisions~\cite{PhysRevD.77.044036,PhysRevD.106.124011,croft2023gravitational,ge2025dynamics}, quasi-circular mergers~\cite{PhysRevD.96.104058,bezares2017final,palenzuela2017gravitational, bezares2022gravitational,evstafyeva2024gravitational,evstafyeva2026lessonsbinarydynamicsinspiralling}, and eccentric binaries~\cite{sanchis2020synchronized,Palloni_2025}. In particular, most studies have focused on computing the  gravitational-wave emission from bosonic star binaries, revealing rich phenomenology and highlighting both similarities and differences with respect to binary black hole and neutron star mergers.

In order to obtain accurate gravitational-waveforms, it is essential to construct appropriate initial data for compact binary systems. For boson and Proca star binaries, however, this task is particularly challenging. In most studies, binary configurations are generated by superposing isolated boson star solutions~\cite{Bezares_2018,PhysRevD.99.024017,PhysRevD.105.104057,Croft_2023}, sometimes supplemented by prescriptions aimed at reducing constraint violations~\cite{PhysRevD.99.044046,Helfer_2022,evstafyeva2023unequal}. Although such constraint-violating initial data provide a convenient starting point for dynamical evolutions, they generically lead to residual violations of Einstein constraint equations and spurious transient dynamics~\cite{PhysRevD.109.044058}. These effects become increasingly relevant for compact configurations and small initial separations, potentially contaminating the early stages of the evolution and the associated gravitational-wave signal. Only recently have fully constraint-satisfying approaches for boson star binaries begun to be developed~\cite{Aurrekoetxea_2023,PhysRevD.107.124018,
PhysRevD.108.124015,PhysRevD.109.044058,aurrekoetxea2025grtresnaopensourcecodesolve}.  

A major computational challenge to build suitable initial data lies in the proper solution of the elliptic constraint equations. Commonly used formulations of the constraint system can suffer from mathematical non-uniqueness when the configuration reaches high compactness. This issue has been investigated in~\cite{PhysRevD.67.044022,doi:10.1142/S0219891605000518} within the framework of the \textit{eXtended Conformal Thin Sandwich} (XCTS) formalism, which extends the \textit{Conformal Thin Sandwich} (CTS) decomposition~\cite{PhysRevLett.82.1350} by supplementing it with an additional elliptic equation for an auxiliary vector field. Early evidence of non-uniqueness in the XCTS elliptic system was first reported in~\cite{PhysRevLett.95.091101}. Further analysis from~\cite{PhysRevD.75.044009} suggested that this behavior is related to the presence of an inappropriate sign in the equation of the conformal factor, which spoils the application of the maximum principle and, as a consequence, does not guaranty the local uniqueness of the solution.

In this work, we develop a fully constrained framework for the construction of initial data for boson star binaries and other non-trivial topologies.
Specifically, our goals are twofold: first, we address the issue of non-uniqueness in XCTS-like elliptic systems in fully three-dimensional settings, focusing on the \textit{eXtended Conformal Flatness Condition} (XCFC)~\cite{cordero-carrion2008}. Second, we use the XCFC formalism to construct fully constraint-satisfying initial data for the aforementioned configurations. The XCFC formulation modifies the \textit{Conformal Flatness Condition} (CFC) scheme~\cite{PhysRevD.54.1317,doi:10.1142/S0218271808011997}, which already contains the relevant elliptic structure of the \textit{Fully Constrained Formalism} (FCF)~\cite{PhysRevD.70.104007,PhysRevD.77.084007}. By construction, XCFC restores a hierarchical elliptic structure with favorable sign properties, allowing the application of maximum-principle arguments to infer the local
uniqueness of the solutions.

This paper is organized as follows. In Sec.~\ref{sec:framework} we briefly review the FCF and CFC formalisms, with an emphasis on the existence and uniqueness properties, and introduce the XCFC framework. Sec.~\ref{sec:code properties} summarizes the main features of the numerical code we have developed to implement the XCFC formalism. In Sec.~\ref{sec:toy_models} we apply our solver to obtain constraint-satisfying initial data for two toy models, namely a Gaussian- and toroidal-like profiles, discussing the stability of the code and providing a convergence test.  Finally, Sec.~\ref{sec:binary_boson_stars} presents the main goal of our work, the construction of initial data for  equal-mass, non-spinning boson star binaries along with a comparison with the data obtained using superposition approaches. Additional details on how numerical choices affect the convergence properties of our code are provided in the Appendix. Throughout this manuscript, we use the signature $\left(-,+,+,+\right)$ for the spacetime metric, and units in which
$c=G=\hbar=1$. Greek indices run from $0$ to $3$, whereas Latin ones run from $1$ to $3$ only.

\section{Framework}
\label{sec:framework}
\subsection{The Fully Constrained Formalism and the Conformal Flatness Condition}
In an asymptotically flat spacetime, one can perform a $3+1$ decomposition of the four-dimensional spacetime metric $g_{\mu\nu}$ on spacelike hypersurfaces $\Sigma_t$, where the data on each hypersurface are given by the pair $(\gamma_{ij},K^{ij})$, with $\gamma_{ij}$ the projected three-dimensional spatial metric onto the hypersurface, and $K^{ij}$  the extrinsic curvature tensor. FCF~\cite{PhysRevD.70.104007} assumes a conformal decomposition of the $3+1$ fields as
\begin{equation}\label{eq:3+1 decomposition}
    \gamma_{ij}=\psi^4\tilde{\gamma}_{ij},\quad\quad K_{ij}=\psi^{4}\tilde{A}_{ij}+\frac{1}{3}K\gamma_{ij},
\end{equation}
where $K=\gamma^{ij}K_{ij}$, $\tilde{\gamma}\mathrel{:=}\det\tilde{\gamma}_{ij}$ and $\tilde{A}^{ij}$ is expressed in terms of the lapse function $N$ and the shift vector $\beta^i$ as
\begin{equation}\label{eq:A decomposition}
    \tilde{A}^{ij}=\frac{1}{2N}\left(\tilde{D}^i\beta^j+\tilde{D}^j\beta^i-\frac{2}{3}\tilde{D}_k\beta^k\tilde{\gamma}^{ij}+\partial_t\tilde{\gamma}^{ij}\right),
\end{equation}
where $\tilde{D}_i$ is the Levi-Civita connection associated with $\tilde{\gamma}_{ij}$. 
Following~\cite{PhysRevD.70.104007}, it is further assumed that the physical metric $\gamma_{ij}$ coincides with a fiducial, time-independent flat metric $f_{ij}$ at spatial infinity, so that  $\tilde{\gamma}=f$ and consequently $\psi=(\gamma/f)^{1/12}$. The deviation (not necessarily small) of the conformal metric from the flat fiducial metric is then defined as
\begin{equation}\label{eq:def_h}
    h^{ij}\mathrel{:=}\tilde{\gamma}^{ij}-f^{ij}.
\end{equation}
The gauge conditions used in~\cite{PhysRevD.70.104007} are the maximal slicing, $K=0$, and the generalized Dirac gauge, $\mathcal{D}_k\tilde{\gamma}^{ki}=0$, where $\mathcal{D}$ stands for the covariant derivative associated with the fiducial flat metric $f_{ij}$. Under these choices, the Einstein equations split into an elliptic-hyperbolic system, where the elliptic sector includes the Hamiltonian and momentum constraint equations.

In particular, when the condition $h^{ij}=0$ is imposed, the resulting spatial three-metric is conformally flat and CFC is then recovered. This shows that the FCF is a natural generalization of the CFC approximation. Under this assumption, the elliptic subsystem of the FCF reduces for CFC to
\begin{equation}\label{eq:Ham_FCF}
    \Delta \psi=-2\pi\psi^{-1}\left[E^*+\frac{\psi^6K_{ij}K^{ij}}{16\pi}\right],
\end{equation}
\begin{equation}\label{eq:dtK_FCF}
    \Delta (N\psi)=2\pi N\psi^{-1}\left[E^*+2S^*+\frac{7\psi^6K^{ij}K_{ij}}{16\pi}\right],
\end{equation}
\begin{equation}\label{eq:Mom_FCF}
    \Delta \beta^i+\frac{1}{3}\mathcal{D}^i
    \mathcal{D}_j\beta^j=16\pi N\psi^{-2}\left(S^*\right)^i+2\psi^{10}K^{ij}\mathcal{D}_j\frac{N}{\psi^6}.
\end{equation}
The starred quantities in the above equations are conformally rescaled matter quantities defined as~\cite{York1979Sources}
\begin{equation}
    E^*\mathrel{:=} \sqrt{\gamma/f}\ E=\psi^6E,
\end{equation}
\begin{equation}
    S^*\mathrel{:=} \sqrt{\gamma/f}\ S=\psi^6S,
\end{equation}
\begin{equation}
    (S^*)_i\mathrel{:=} \sqrt{\gamma/f}\ S_i=\psi^6S_i,
\end{equation}
where $E\mathrel{:=} T_{\mu\nu}n^{\mu}n^{\nu}$, $S^i\mathrel{:=} -\gamma^{i\mu}T_{\mu\nu}n^{\nu}$ and $S\mathrel{:=} -\gamma^{ij}S_{ij}$, with $S_{ij}\mathrel{:=}T_{\mu\nu}\gamma^{\mu}_i\gamma^{\nu}_j$ are, respectively, the energy density, momentum density and the trace of the stress-energy tensor, all measured by the Eulerian observer (meaning an observer with four-velocity $n^{\mu}$).

Eqs.~(\ref{eq:Ham_FCF}) and~(\ref{eq:dtK_FCF}) inherit the local non-uniqueness issues already present in the elliptic system of constraint equations within the FCF, a problem that was addressed in previous references (e.g.~\cite{PhysRevLett.95.091101,PhysRevD.75.044009, Rinne_2008}) and will be explained in more detail in the next subsections.

\subsection{Uniqueness and existence problems}\label{subsect::uniqueness}

Systems of elliptic equations can accept multiple solutions, according, for example, to different boundary conditions. However, one should single out the one that is consistent to the physical problem we want to model. To achieve this, the free quantities of the elliptic equations must be chosen appropriately, without over-restricting the system, guaranteeing the existence of a solution: this is known as the $existence$ problem. On the other hand, fewer restrictions could provide a non-unique solution: this is the $uniqueness$ problem.

To further discuss the uniqueness of the solution for an elliptic equation, we can address the following problem in a regular domain $\Omega \subset \mathbb{R}^n$,
\begin{equation}
    \Delta u + f(u) = g \quad \mathrm{in}~\Omega,
\end{equation}
where $g$ is a known function independent of $u$, $f(u)$ is a non-increasing function, and fixed boundary conditions are prescribed on $\partial\Omega$.  

Suppose that $u$ and $v$ are two solutions of the same equation with identical boundary conditions.  
Setting $w = u - v$, we have $w = 0$ on $\partial\Omega$.  
Subtracting the two equations produces a new elliptic equation for $w$, which contains the nonlinear term $f(u) - f(v)$.  
Since $f$ is non-increasing, this term always has the opposite sign of $w$, implying that the Laplacian of $w$ is non-negative in the region where $w>0$. By applying the maximum principle, we deduce that $w$ cannot attain a positive maximum inside $\Omega$, and therefore $w \le 0$ throughout the domain. Repeating the same argument after interchanging $u$ and $v$ gives $w \ge 0$, hence $w \equiv 0$ in $\Omega$. This proves that the solution of the elliptic equation is unique.

If we assume in particular that $f(u) = h(x)\,u^p$, where $h$ is a function independent of $u$, the same argument shows that the maximum principle guarantees local uniqueness of the solution whenever the sign of the exponent $p$ is opposite to that of $h(x)$. This sign condition ensures that $f(u)$ remains non-increasing, which is exactly the hypothesis required for the argument to hold~\cite{York1979Sources,taylor1996partial, protter1967maximum, evans1998partial}.

In CFC one is dealing with a system of elliptic equations, Eqs.~({\ref{eq:Ham_FCF}}) to (\ref{eq:Mom_FCF}), assessing whether they exhibit an appropriate signs for the application of the maximum principle, which is crucial for  understanding the uniqueness problem. Once all quantities are expressed in terms of the lapse and the shift, the presence of an inappropriate sign becomes apparent, which, in principle, might yield non-uniqueness, leading to convergence to undesirable solutions.

\subsection{The eXtended  Conformal Flatness Condition}

In the XCFC, introduced in~\cite{cordero-carrion2008}, the main idea is to generalize to the FCF framework the situation encountered in spherical symmetry, where the only non-vanishing component of the conformal extrinsic curvature, $\psi^6 K^r_{\ r}$, can be computed explicitly. This decouples the equation for $\psi$ from the rest of the elliptic system, thereby overcoming the issue of local non-uniqueness and ensuring convergence of the solver toward a physically correct solution.

In order to achieve this goal, two different conformal decompositions of the extrinsic curvature must be introduced, involving two distinct conformal rescalings and two different decompositions of the traceless part into longitudinal and transverse components.  
Assuming the maximal slicing condition $K=0$, a generic conformal decomposition can be written as
\begin{equation}
    K^{ij} = \psi^{\zeta - 8} \left(A^{(\zeta)}\right)^{ij} 
    \mathrel{:=} \psi^{\zeta - 8} \left[\frac{1}{\sigma} \left(LX\right)^{ij} + A^{ij}_{\rm TT}\right],
\end{equation}
where $\zeta$ is a free parameter and $\sigma$ a freely specifiable function.  Here, $A^{ij}_{\rm TT}$ denotes the transverse–traceless part (with respect to a conformally flat metric), while the vector $X^i$ represents the longitudinal component of $\left(A^{(\zeta)}\right)^{ij}$. The operator $L$ is the conformal Killing operator, defined as
\begin{equation}
    \left(LX\right)^{ij} \mathrel{:=} \mathcal{D}^i X^j + \mathcal{D}^j X^i - \frac{2}{3} f^{ij} \mathcal{D}_k X^k.
\end{equation}
This is essentially the same decomposition introduced in Eqs.~(\ref{eq:3+1 decomposition}) and (\ref{eq:A decomposition}), but now with $\zeta = 4$ and $\sigma = 2N$, corresponding to a conformal thin sandwich (CTS)-like decomposition~\cite{PhysRevLett.82.1350} of the traceless part. 
This choice allows the vector $X^i$ to be identified with the shift vector, and $A^{ij}_{\rm TT}$ to be expressed in terms of the time derivative of the conformal metric:
\begin{equation}\label{eq:CTS decomposition}
    K^{ij} = \psi^{-4} \tilde{A}^{ij}, 
    \quad\quad 
    \tilde{A}^{ij} = \frac{1}{2N} \left(L\beta\right)^{ij}.
\end{equation}
The second decomposition, corresponding to  conformal transverse traceless (CTT), assuming $\zeta=-2$ and  $\sigma=1$ is
\begin{equation}\label{eq:CTT decomposition}
    K^{ij} = \psi^{-10} \hat{A}^{ij}, 
    \quad\quad 
    \hat{A}^{ij} =  \left(LX\right)^{ij}+\hat{A}^{ij}_{\rm TT}.
\end{equation}
Using the second decomposition, the CFC momentum constraint can be written as
\begin{equation}\label{eq:momentum constraint in CTT}
    \mathcal{D}_j\hat{A}^{ij}=8\pi\psi^{10}S^i=8\pi\psi^6f^{ij}S_j=8\pi f^{ij}S_j^*.
\end{equation}
The connection between the two decompositions is given by
\begin{equation}\label{eq:CTT-CTS}
    \hat{A}^{ij}=\psi^{10}K^{ij}=\psi^6\tilde{A}^{ij}.
\end{equation}
Under the assumption that the transverse traceless component is smaller in amplitude than the nonconformal part $h^{ij}$ of the spatial metric (an assumption consistent with the accuracy of the CFC approximation, as shown in the Appendix of~\cite{cordero-carrion2008}) we can consider that
\begin{equation}\label{eq:def hatA}
    \hat{A}^{ij}\approx(LX)^{ij}=\mathcal{D}^iX^j+\mathcal{D}^jX^i-\frac{2}{3}\mathcal{D}_kX^kf^{ij}.
\end{equation}
From Eqs.~(\ref{eq:CTT decomposition}) and (\ref{eq:momentum constraint in CTT}), one can derive an elliptic equation for $X^i$:
\begin{equation}\label{eq:X}
    \Delta X^i+\frac{1}{3}\mathcal{D}^i\mathcal{D}_jX^j=8\pi f^{ij}S^*_j.
\end{equation}
The elliptic equation for the conformal factor could be rewritten in terms of the conformally rescaled quantities and $\hat{A}_{ij}$ as
\begin{equation}\label{eq:psi}
    \Delta\psi=-2\pi\psi^{-1}E^*-\psi^{-7}\frac{f_{il}f_{jm}\hat{A}^{lm}\hat{A}^{ij}}{8}.
\end{equation}
Once the conformal factor is recovered, the non-rescaled stress-energy tensor components can be recovered. The elliptic equation for $N\psi$ can be solved, obtaining the lapse function $N$:
\begin{equation}\label{eq:Npsi}
    \Delta(N\psi)=2\pi N\psi^{-1}\left(E^*+2S^*\right)+N\psi^{-7}\frac{7f_{il}f_{jm}\hat{A}^{lm}\hat{A}^{ij}}{8}.
\end{equation}
These equations ensure the suitable sign required for the maximum principle to be satisfied. Finally, by taking the divergence of Eq.~(\ref{eq:CTS decomposition}) and using Eq.~(\ref{eq:CTT-CTS}) one can derive an elliptic equation for the shift vector,
\begin{equation}\label{eq:beta}
    \Delta\beta^i+\frac{1}{3}\mathcal{D}^i\left(\mathcal{D}_j\beta^j\right)=\mathcal{D}_j\left(2N\psi^{-6}\hat{A}^{ij}\right).
\end{equation}
With this reformulation of the CFC equations the sign of the exponents of conformal factor and lapse become consistent with the maximum principle for scalar elliptic equations, ensuring linearization stability of the system. The main drawback of this approach is the introduction of an additional elliptic  equation for the vector field $X^i$. Although this procedure does not guaranty the global uniqueness of the solution, it is a sufficient condition for  local uniqueness. A similar strategy can be applied to the elliptic sector of the FCF~\cite{cordero-carrion2008}. Using the FCF is beyond the scope of this work. However, as we show below, relevant improvements in comparison with other approaches already become apparent within the CFC context.

\begin{table*}[t]
\centering
\caption{Computational time required for convergence for the different configurations considered in this work, highlighting the number of processors employed along each spatial dimension, the grid resolution, and the relaxation, convergence, and outer-iteration thresholds. The models shown include the Gaussian (G) and toroidal (T) scalar-field profiles analyzed in Secs.~\ref{subsec:gaussian_scalar_field_profile} and \ref{subsec:toroidal_profile}, respectively; binary boson stars without initial boost with $\phi_0=5\times10^{-3}$ (BBS1), $\phi_0=3\times10^{-2}$ (BBS2), and $\phi_0=4.5\times10^{-2}$ (BBS3), discussed in Sec.~\ref{subsec:BBS_noboost}; and boosted binary boson stars with $\phi_0=2\times10^{-2}$ and $v=0.01$ (bBBS1), and $v=0.20$ (bBBS2), discussed in Sec.~\ref{subsec:BBS_boost}. All results presented in this work were obtained using the MareNostrum~5 supercomputer hosted at the Barcelona Supercomputing Center.}
\label{tab:time}

\begin{tabular}{lcccccc}
\toprule
Model & $N_p^3$ & Relaxation & Convergence & Outer-iteration & MPI processors & Time [h] \\
\midrule
G & $50^3$ & $10^{-15}$ & $10^{-15}$ & - & $2$ & $3$\\
G & $100^3$ & $10^{-15}$ & $10^{-15}$ & - & $2$ & $20$\\
T & $200^3$ & $10^{-13}$ & $10^{-13}$ & - & $5$ & $18$ \\
T & $250^3$ & $10^{-13}$ & $10^{-13}$ & - & $5$ & $48$ \\
BBS1 & $200^3$ & $10^{-10}$ & $10^{-10}$ & - & $5$ & $<1$ \\
BBS1 & $250^3$ & $10^{-10}$ & $10^{-10}$ & - & $5$ & $1.5$\\
BBS2 & $250^3$ & $10^{-10}$ & $10^{-10}$ & - & $5$ & $12$\\
BBS3 & $200^3$ & $10^{-10}$ & $10^{-10}$ & - & $5$ & $18$\\
BBS3 & $250^3$ & $10^{-10}$ & $10^{-10}$ & - & $5$ & $47$\\
bBBS1 & $200^3$ & $10^{-10}$ & $10^{-10}$ & $10^{-10}$ & $5$ & $9$\\
bBBS1 & $250^3$ & $10^{-10}$ & $10^{-10}$ & $10^{-10}$ & $5$ & $24$\\
bBBS2 & $200^3$ & $10^{-10}$ & $10^{-10}$ & $10^{-10}$ & $5$ & $22$\\
bBBS2 & $250^3$ & $10^{-10}$ & $10^{-10}$ & $10^{-10}$ & $5$ & $59$\\
\bottomrule
\end{tabular}

\end{table*}

\section{Basic code features}
\label{sec:code properties}

The numerical code we have developed to solve the XCFC system of elliptic equations, named \textsc{Incipit}, is structured in a hierarchical way, on which local uniqueness strongly relies. It extends the framework previously adopted in~\cite{sanchis2019nonlinear,DiGiovanni:2022mkn} for the construction of initial data describing clouds of spinning scalar and Proca fields, and rotating neutron stars, respectively. In the code, 
after computing the non–conformally rescaled components of the stress–energy tensor (either analytically or using external codes; see below), we provide an initial guess for the metric functions, aiming to converge toward a physically consistent solution compatible with the chosen matter configuration.
The overall solving procedure can be summarized as follows:
\begin{enumerate}
    \item With the conformally rescaled quantities, we solve Eq.~(\ref{eq:X}) for $X^i$ and thus, through Eq.~(\ref{eq:def hatA}), $\hat{A}^{ij}$. 
    \item Using Eq.~(\ref{eq:psi}) we compute the conformal factor.
    \item We then apply the maximum principle in Eq.~(\ref{eq:Npsi}), to guarantee existence and uniqueness of the solution. Solving this equation for $N\psi$, we  compute the lapse function.
    \item Finally, as the source in Eq.~(\ref{eq:beta}) is fully known, we solve this equation for $\beta^i$.
\end{enumerate}
Once an initial solution for the metric functions is obtained, the code updates their values and recomputes the conformally rescaled components of the stress-energy tensor. This iterative procedure is repeated until convergence is achieved, namely the overall residual falls below a prescribed convergence threshold.

The code is based on a uniform grid and can be used in full three dimensions without assuming any symmetries, employing Cartesian coordinates. Moreover, depending on the physical problem under consideration, it is also possible to change the coordinate system, for instance, to spherical coordinates, as well as to select the dimensionality of the code accordingly. Additionally, the code has an MPI-based parallelization scheme, which is crucial for dealing with three-dimensional configurations.

The finite-difference discretization of the elliptic equations is fourth-order accurate, and several types of boundary conditions are implemented, namely Dirichlet, Neumann, Robin, and generalized Robin. In this work, only Dirichlet and Robin boundary conditions are used, as they provide sufficient accuracy for our purposes. The solver adopted for our results is a modified version of the optimized \textit{Scheduled Relaxation Jacobi} (SRJ) method~\cite{ADSUARA2016369,ADSUARA2017446}, with relaxation thresholds adjusted to the specific scenario under investigation (see specific subsections below for more details); in our implementation, the weights of the SRJ solver, $w_i$, are constrained to $w_i\in [0.5,1.5]$, preventing possible divergence associated with excessively large weights. The results presented in this paper, along with the associated CPU timing estimates, were obtained using the MareNostrum~5 supercomputer, hosted at the Barcelona Supercomputing Center. A detailed description of the system specs can be found in~\cite{banchelli2025introducingmarenostrum5europeanpreexascale}.

\section{Results using toy models}
\label{sec:toy_models}

We start by assessing our code on results obtained using two toy models: a massive (complex) scalar field with a Gaussian-like profile and a massive (complex) scalar field with a toroidal profile. For both cases, we employ a three-dimensional setup in Cartesian coordinates with a non-staggered grid, using an MPI parallelization scheme with $2$ processors per spatial direction in the first case and $5$ in the second. In order to consistently evaluate the numerical error introduced by our solver and to perform reliable convergence tests, we set  convergence and relaxation thresholds of $10^{-15}$ and $10^{-13}$, respectively, for both toy models. Those very strict thresholds ensure
that any spurious effects are completely suppressed and convergence is achieved.
As a consequence, the computational time required by the code to achieve convergence 
is relatively long. Table~\ref{tab:time} summarizes the computational time required to achieve convergence for the different configurations investigated in this work. We assume the initial guess for the metric functions to be the Minkowskian flat metric, so as not to bias the convergence toward any particular solution. In the Gaussian profile case we employ Dirichlet-type boundary conditions for the parallel domains and Robin-type for the outer boundaries of the grid. For the toroidal-like profile, Dirichlet boundary conditions are instead imposed on the outer boundaries of the grid. To assess the magnitude of the numerical errors, we consider the Hamiltonian and momentum constraints in their general forms, respectively:
\begin{eqnarray}
    \mathcal{H} &=& R + K^2 - K_{ij} K^{ij} - 16\pi E,
    \\
    \mathcal{M}^i &=& D_j\left(K^{ij}-\gamma^{ij}K\right) - 8\pi S^i,
\end{eqnarray}
where $R$ denotes the Ricci scalar and $D_j$ the covariant derivative associated with $\gamma_{ij}$. Under the assumptions of maximal slicing and asymptotic flatness, these equations reduce to the formulation given in Eqs.~(\ref{eq:Ham_FCF}) and (\ref{eq:Mom_FCF}). For the first toy model, since the matter content is spherically symmetric, the constraint equations further correspond to the XCFC formulation (Eqs.~(\ref{eq:psi}) and (\ref{eq:momentum constraint in CTT}), respectively), which in this case is no longer an approximation.

\subsection{Gaussian scalar field profile}
\label{subsec:gaussian_scalar_field_profile}

\begin{figure}[t] 
    \centering
    \includegraphics[width=0.45\textwidth]{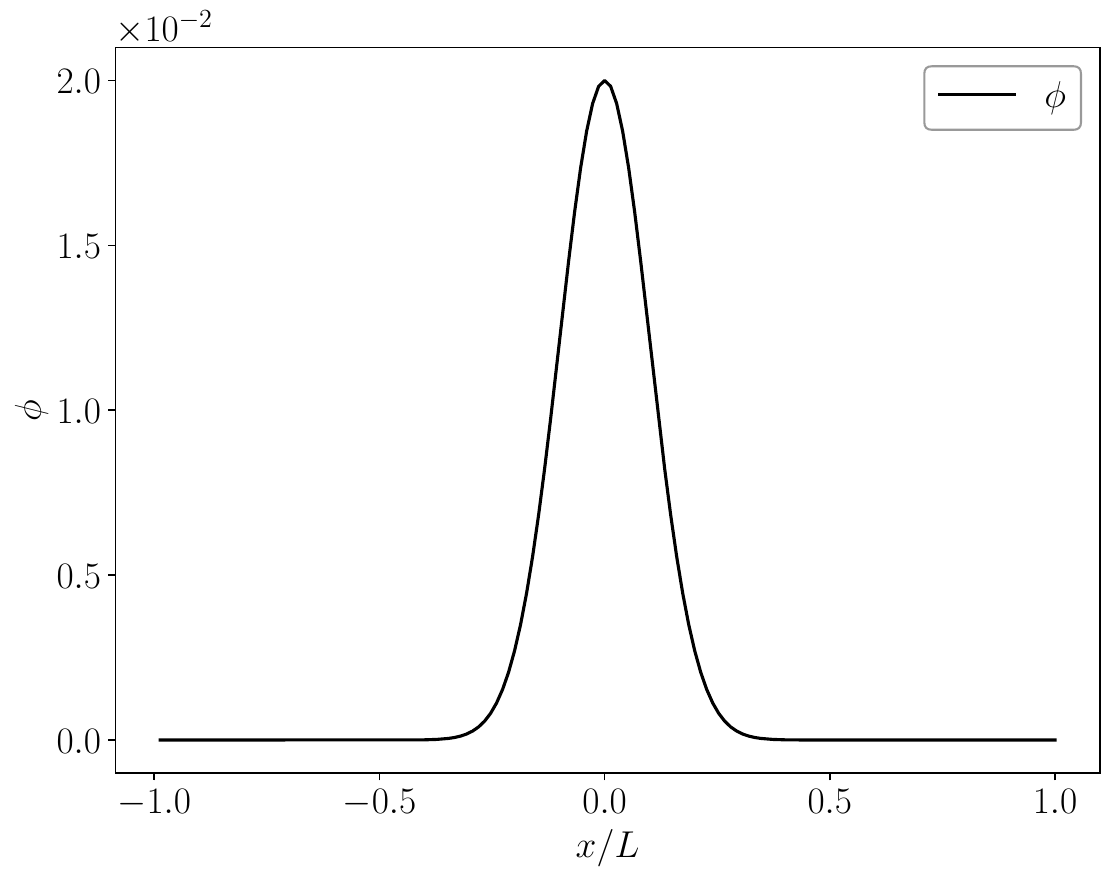} 
    \includegraphics[width=0.46\textwidth]{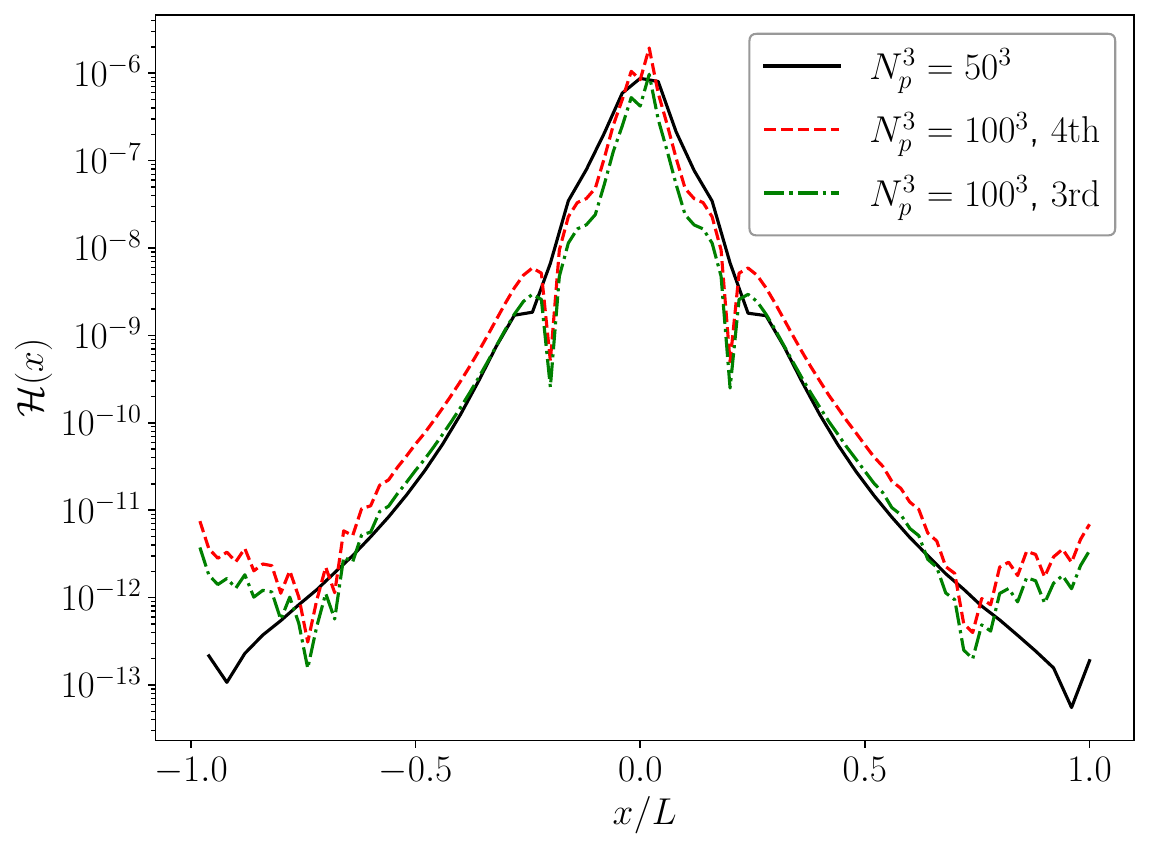}
    \caption{\textit{Top:} Gaussian-like profile of a massive complex scalar field with $A=2\times10^{-2}$ and $\sigma=2$ along the $x$-axis. \textit{Bottom:}  Convergence test for the Hamiltonian constraint violation along the $x$-axis for different resolutions, corresponding to $N_p^3=(50^3,100^3)$, shown as solid black and dashed red (green) lines respectively. The higher-resolution results have been rescaled according to fourth-order (third-order) convergence. In both plots the $x$-axis has been  normalized with respect to $L=10\,\sigma$.}
    \label{fig:gaussian_hamiltonian_constraint_convergence}
\end{figure}

The first test we perform involves a massive complex scalar field with a three-dimensional Gaussian profile and a harmonic time dependence,
\begin{equation}
    \Phi(t, \vec{x})=\phi(\vec{x})e^{i\omega t}=Ae^{-( x^2+y^2+z^2)/ 2\sigma^2}e^{ i\omega t},
\end{equation}
where $A$ is the amplitude, $\sigma$ the variance, and $\omega$ the frequency (which we assume equal to $1$ without any loss of generality).  The Gaussian is centered at the origin of the Cartesian system of coordinates.

The associated stress-energy tensor is given by
\begin{eqnarray}
\label{eq:Tmunu_field}
    T_{\mu\nu}&=&\frac{1}{2}\left(\nabla_{\mu}\Phi^*\nabla_{\nu}\Phi+\nabla_{\mu}\Phi\nabla_{\nu}\Phi^*\right)
    \nonumber \\
    &-&\frac{1}{2}g_{\mu\nu}\left(g^{\rho\sigma}\nabla_{\rho}\Phi^*\nabla_{\sigma}\Phi + V\left(|\Phi|^2\right)\right),
\end{eqnarray}
where $\nabla_{\mu}$ stands for the covariant derivative with respect to the spacetime metric $g_{\mu\nu}$, the star denotes complex conjugate variables and $V\left(|\Phi|^2\right)$ is the scalar potential associated with the field. We consider a solitonic potential with no self-interaction (quadratic scalar potential),
\begin{equation}
    V(|\Phi|^2) = \mu^2 |\Phi|^2 = \mu^2 \phi^2,
\end{equation}
where $\mu$ is the mass of the bosonic particle. For simplicity we assume $\mu=1$. For this field, one can define a conjugate momentum $\Pi$,
\begin{equation}
    \Pi=\frac{1}{N}\left(\partial_t-\beta^i\partial_i\right)\Phi.
\end{equation}

Given the parameter $\sigma$, the computational domain is chosen such that its length is an integer multiple of $2\sigma$. The total number of points in the box is $N_p^3$, where $N_p$ is the number of points for each axis. After computing the components of the stress-energy tensor and assuming the Minkowskian metric as the initial guess, we solve the system of equations in the XCFC approximation. As expected analytically, since the system is spherically symmetric, the shift vector $\beta^i$, the longitudinal component of the tensor $X^i$ and, by construction (due to the assumptions made), the momentum constraint violation remains zero throughout the computation.

\begin{figure*}[t] 
    \centering
    \includegraphics[width=0.45\textwidth]{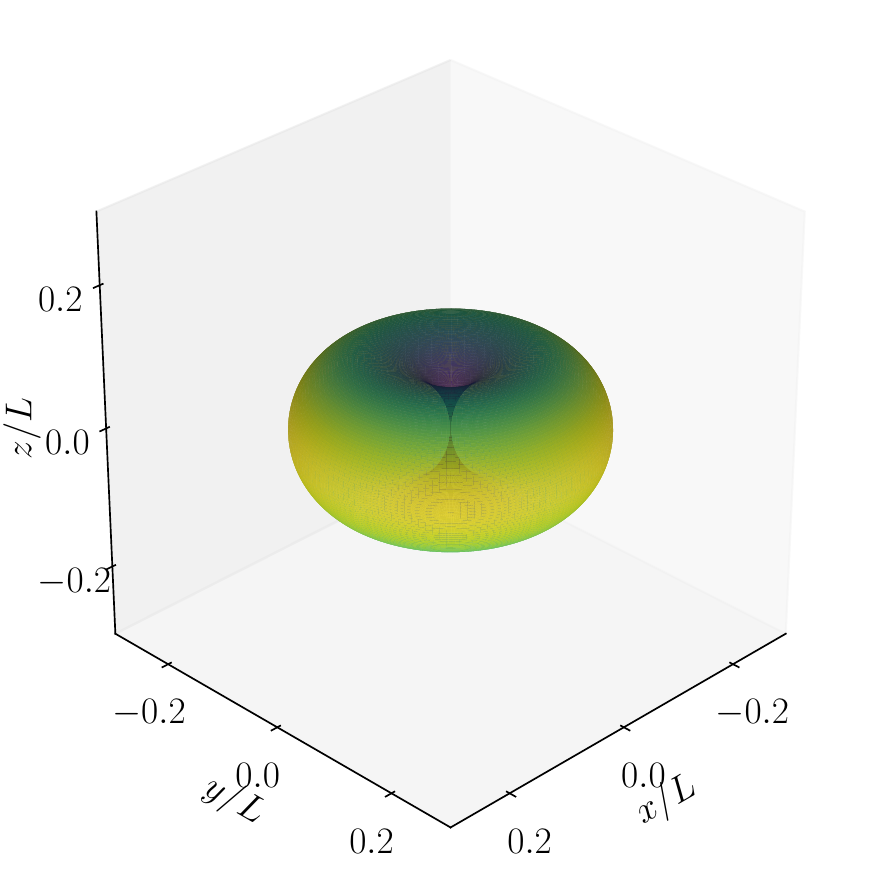} 
    \includegraphics[width=0.45\textwidth]{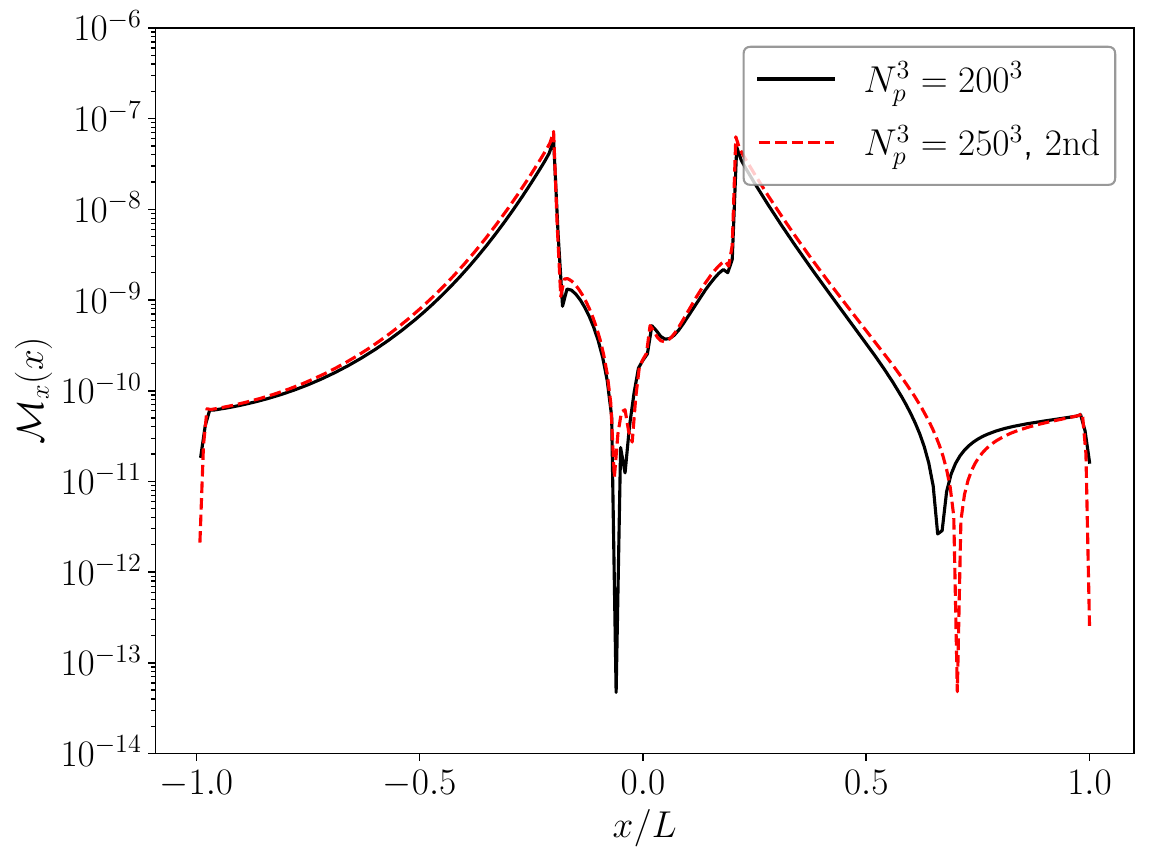} \\
    \includegraphics[width=1\textwidth]{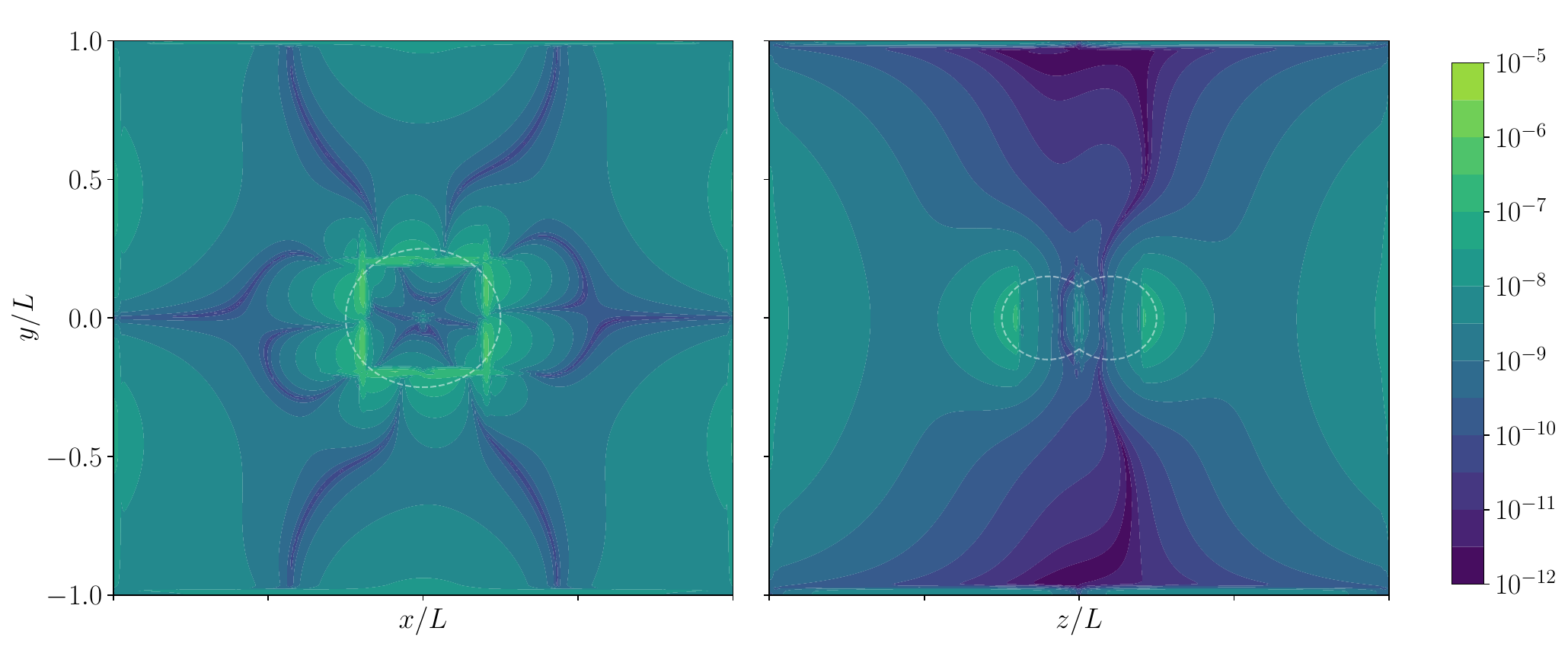} 
    \caption{\textit{Top-left panel:} Toroidal-like profile of a massive complex scalar field with $A=5\times10^{-4}$ (amplitude), $\sigma=2.0$ (variance), $a=2.0$ (radius of the torus),  $m=1.0$ (angular momentum number), and $\omega=1$ (frequency on the equatorial plane). \textit{Top-right panel:} $x-$component of the momentum constraint violation on the $x$-axis for $N_p^3={200^3,250^3}$, solid black and dashed red lines, respectively. The former has been rescaled according to 2nd order convergence. \textit{Bottom panels:} $x$-component of the momentum constraint along the equatorial ($xy$-, on the left) plane and $yz$-plane (on the right) for $N_p^3=250^3$.}
    \label{fig:torus_hamiltonian_constraint_convergence}
\end{figure*}

The top panel of Fig.~\ref{fig:gaussian_hamiltonian_constraint_convergence} shows the profile of the scalar field along the $x$-axis (top) for a configuration with $A = 2 \times 10^{-2}$ and $\sigma = 2$. The computational grid extends over a domain that is $20$ times the variance in length, such that each variable runs from $-L=-10\,\sigma$ to $L=10\,\sigma$  (in the plot we have normalized the grid for simplicity). We chose $N_p^3=100^3$, which corresponds to the highest resolution for the case under examination. A convergence test for the Hamiltonian constraint violation along the $x$-axis is shown in the bottom panel. The resolutions considered are $N_p^3=(50^3,100^3)$, represented by the solid black and dashed red lines, respectively. The latter has been appropriately rescaled according to fourth-order convergence. The dashed green curve represents the rescaled constraint violation for a resolution $N_p^3=100^3$ assuming a convergence order of $3$. We can therefore conclude that the convergence order lies between $3$ and $4$, in good agreement with the theoretically expected discretization order. This result confirms the reliability and accuracy of the adopted numerical scheme.

Convergence may be affected by several factors, including the choice of boundary conditions as well as the convergence and relaxation thresholds. For completeness, we discuss in Appendix~\ref{app:1} the impact of using wrong boundary conditions and different thresholds on the convergence properties for this particular test (and some more details on the Robin boundary conditions implemented).

\subsection{Toroidal scalar field profile}
\label{subsec:toroidal_profile}

The second test we perform consists in a massive complex scalar field with a toroidal topology, which, in cylindrical coordinates is given by:
\begin{eqnarray}
    \Phi(t, \vec{x})&=&\phi(\vec{x})e^{i(m\varphi-\omega t)}\nonumber\\
    &=&A\rho^{|m|}e^{-(\left(\rho-a\right)^2+z^2)/2\sigma^2}e^{i(m\varphi-\omega t)}\,.
\end{eqnarray}
Here, $A$ is the amplitude, $\sigma$ the variance, $m$ is the angular momentum number, $a$ the radius of the torus, and $\omega$ the field frequency. As in the previous case we assume $\omega=1$. For simplicity, we consider that the system rotates with the lowest angular momentum number possible, namely $m=1$. Since we want to test our code for Cartesian coordinates, we also perform the corresponding  coordinate transformation. Moreover, as in the previous toy model, the length of the computational domain is chosen to be an integer multiple of $2\sigma$, with total number of grid points $N_p^3$, where $N_p$ denotes the number of points along each spatial direction.

As in the previous test, we assume a Minkowskian metric as the initial guess and solve the system of equations within the XCFC approximation. For this particular configuration, both the shift vector $\beta^i$ and the longitudinal component of tensor $X^i$ are non-trivial for equilibrium static configurations, as is the asymptotic conformal metric (i.e.\ $h_{ij} \neq 0$; see Eq.~(\ref{eq:def_h})). Nevertheless, this setup is fully consistent with Einstein's equations (as all constraints are fulfilled). Despite not corresponding to a static configuration, the difference between the computed metric variables and those of a static configuration is expected to be very sufficiently small, as shown in~\cite{tbz7-hpcy}.

The top-left panel of Fig.~\ref{fig:torus_hamiltonian_constraint_convergence} shows the profile of the scalar field along the equatorial plane for a resolution of $N_p^3 =250^3$, with parameters $A = 5 \times 10^{-4}$, $\sigma = 2.0$, $a = 2.0$. The computational grid extends over a region that is $20$ times the variance length, such that each spatial axis ranges from $-L = -10\,\sigma$ to $L = 10\,\sigma$. To study the convergence properties of the solution, we consider two resolutions, $ N_p^3 = (200^3, 250^3)$. We note that lower resolutions do not  properly resolve the matter topology of the torus. The top-right plot of Fig.~\ref{fig:torus_hamiltonian_constraint_convergence} shows the convergence test on the $x$-component of the momentum constraint along the $x$-axis (conveniently rescaled by $L$) for the two resolutions. The solid black line corresponds to $N_p^3 = 200^3$ and the dashed red line to $N_p^3 = 250^3$. The line for the coarsest resolution has been rescaled according to second-order convergence, which is slightly smaller than the expected one from the methods used. The low convergence order may be influenced by the numerical resolution, as the values employed might not lie within the asymptotic convergence regime. Increasing the resolution further would not be worthwhile given our current computational resources. For the finest resolution, the model requires nearly two days to reach convergence (see Table~\ref{tab:time}), despite being intended primarily as a toy model, whereas our main focus is on the physical setup of binary boson stars (see next section). In addition, the boundary conditions may also play a role on the lower convergence order achieved: in this setup Dirichlet conditions were adopted for their computational efficiency whereas simulations using Robin boundary conditions did not achieve convergence.

The bottom panels of Fig.~\ref{fig:torus_hamiltonian_constraint_convergence} represent the $x$-component of the momentum constraint violation at the $xy$ and $yz$ planes for a resolution of $N_p^3=250^3$. The closed dashed white curves represent the boundary of the $3\sigma$ outer surface of the toroidal scalar field, which coincides with the boundaries of the central parallel domain, where the error reaches its highest values (resulting in a square-like structure in the $xy$ plane). Since the error remains within the range $10^{-12}$--$10^{-5}$, we can conclude that our code is capable of accurately resolving this non-spherically symmetric matter topology.

\section{Binary Boson star initial data}
\label{sec:binary_boson_stars}

\begin{figure*}[t] 
    \centering
    \includegraphics[width=0.45\textwidth]{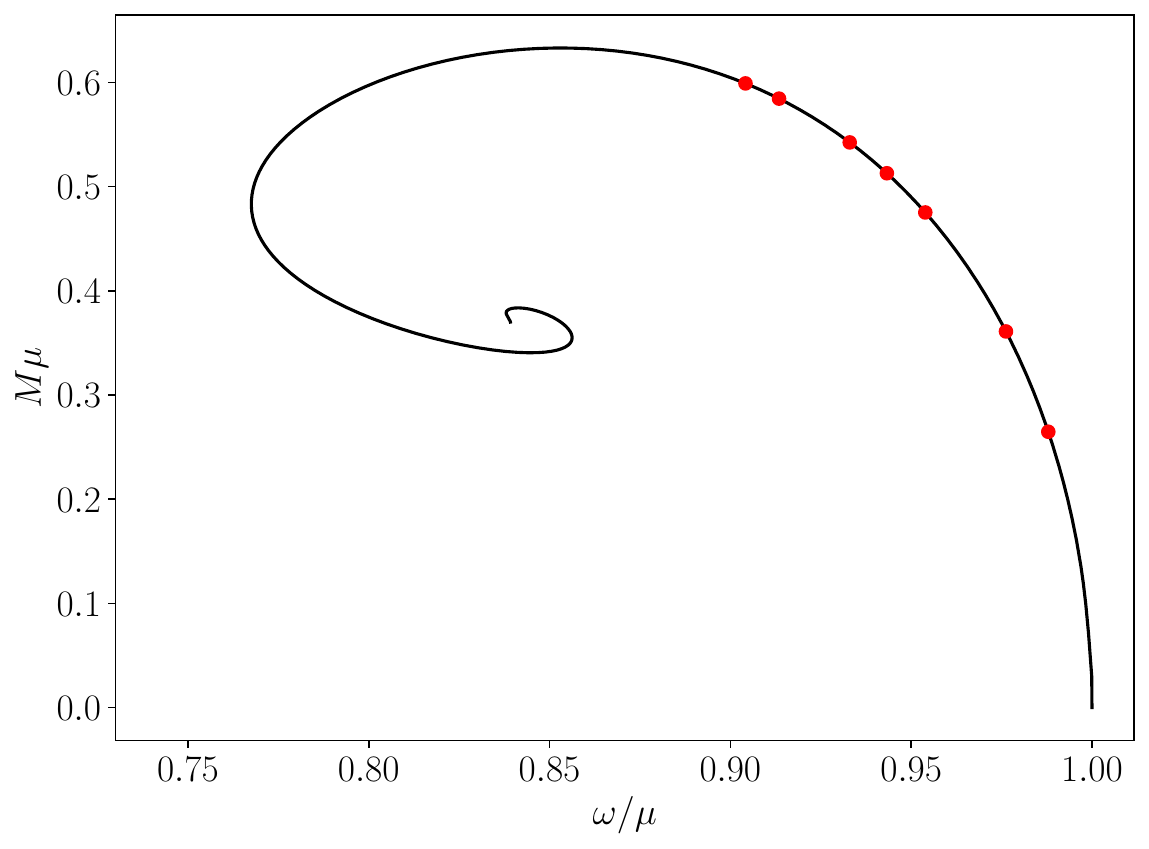} 
    \includegraphics[width=0.45\textwidth]{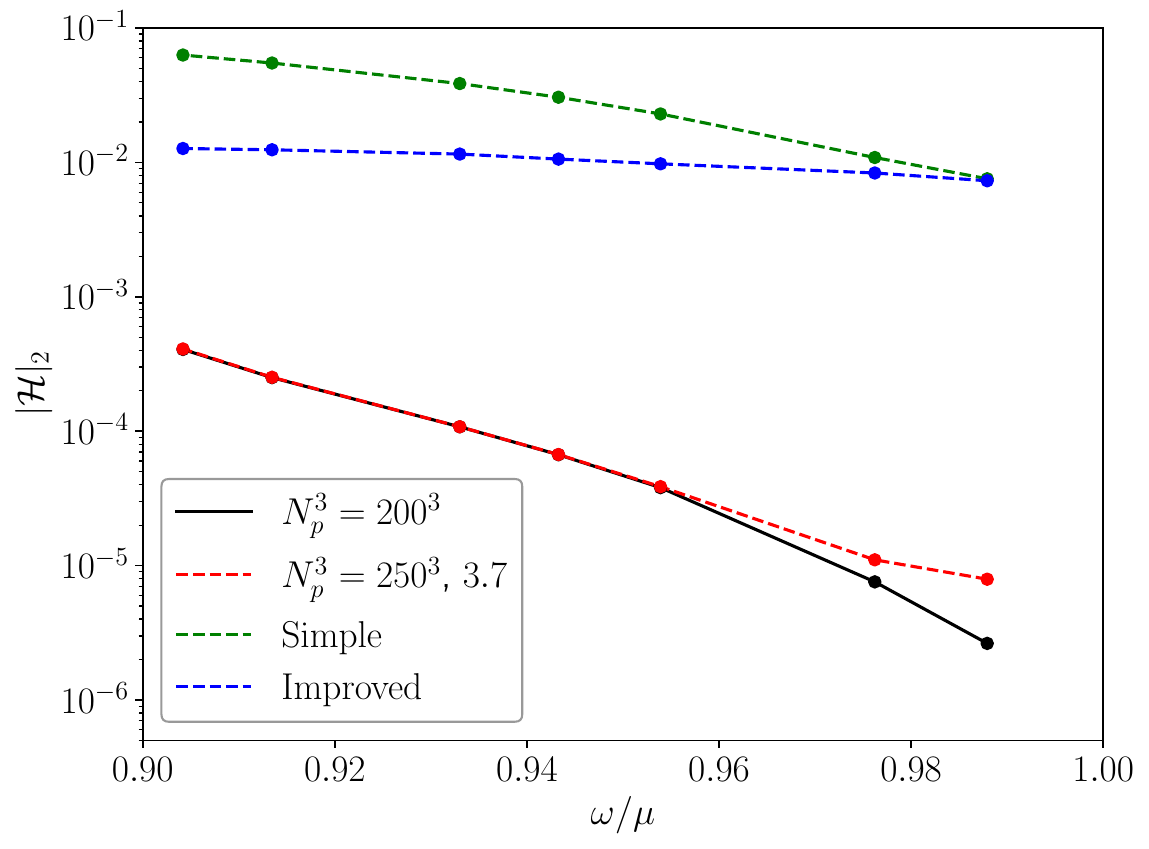}\\
    \includegraphics[width=0.45\textwidth]{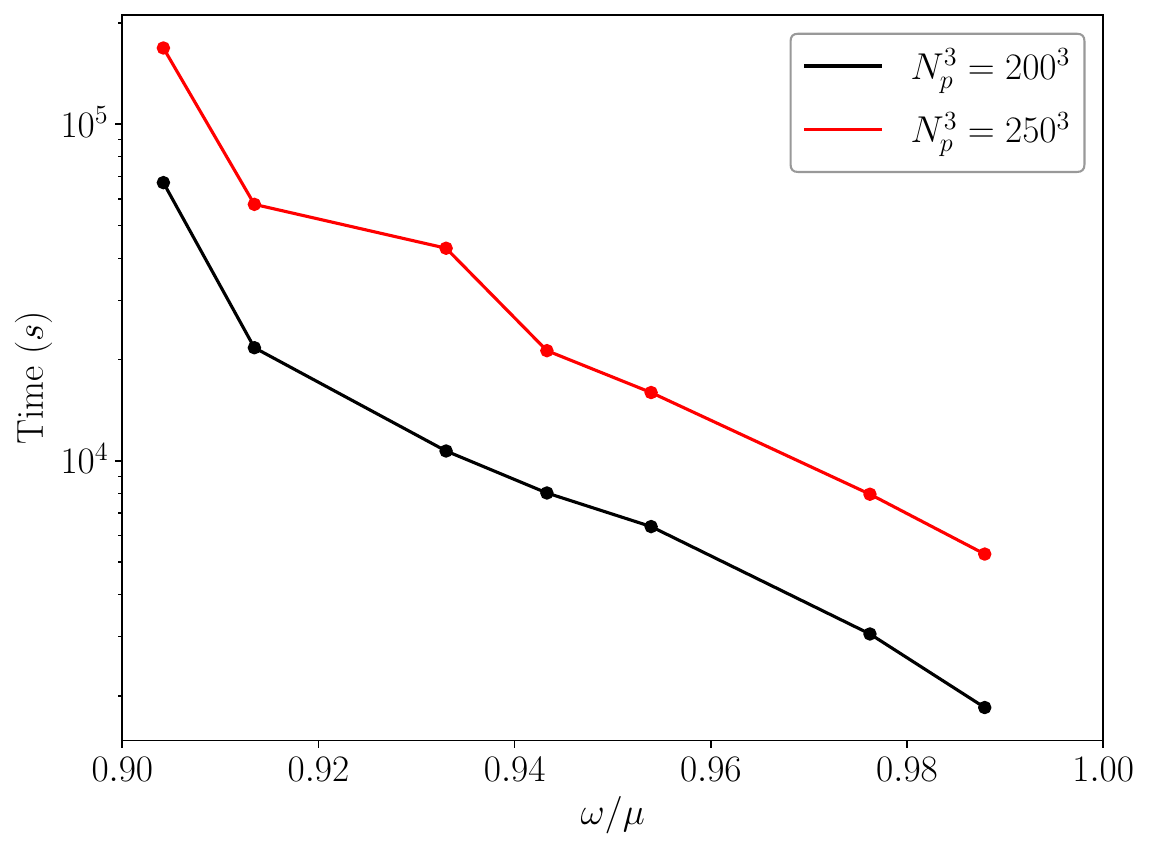} 
    \includegraphics[width=0.45\textwidth]{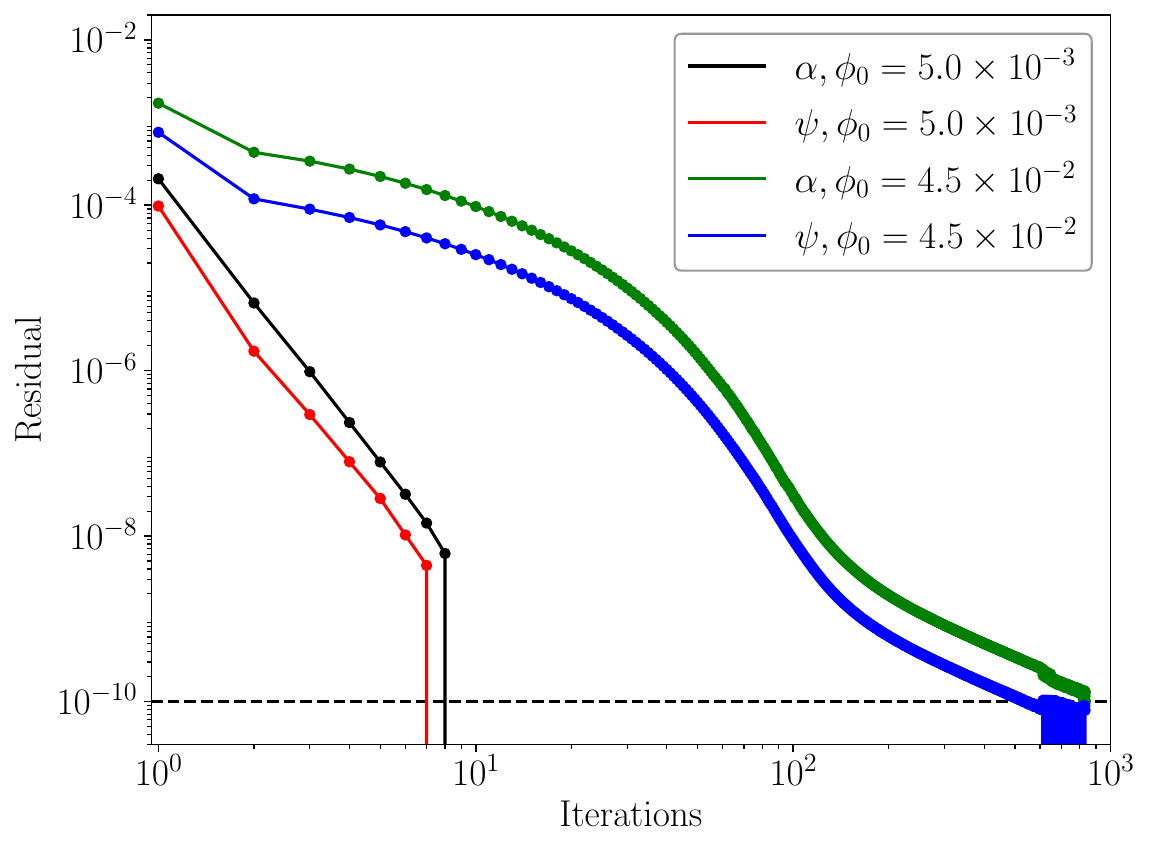}
    \caption{\textit{Top-left panel:} Mass--frequency relation (solid black line) for the boson-star models employed in this study. The red dots indicate the specific models used, all belonging to the stable branch (at the right of the maximum mass, $M\mu \sim 0.63$, corresponding to frequencies greater than $\omega/\mu \sim 0.88$). \textit{Top-right panel:} $L_2$-norm of the Hamiltonian constraint violation at resolutions $N_p^3 = 200^3$ (solid black line) and $N_p^3 = 250^3$ (dashed red line, rescaled according to a convergence order of $3.7$), compared with those obtained from simple superposition (dashed green line) and improved superposition (dashed blue line), both computed at a resolution of $N_p^3 = 250^3$.
    \textit{Bottom-left panel:} Computational time for convergence (in seconds) versus boson star's frequency.
    \textit{Bottom-right panel:} Number of iterations required to achieve convergence for the less (most) compact binary $\phi_0 = 5.0 \times 10^{-3}$ ($\phi_0 = 4.5 \times 10^{-2}$) when solving the lapse and conformal factor.
}
    \label{fig:BBS_all}
\end{figure*}

\begin{figure*}[t] 
    \centering
    \includegraphics[width=1.025\textwidth]{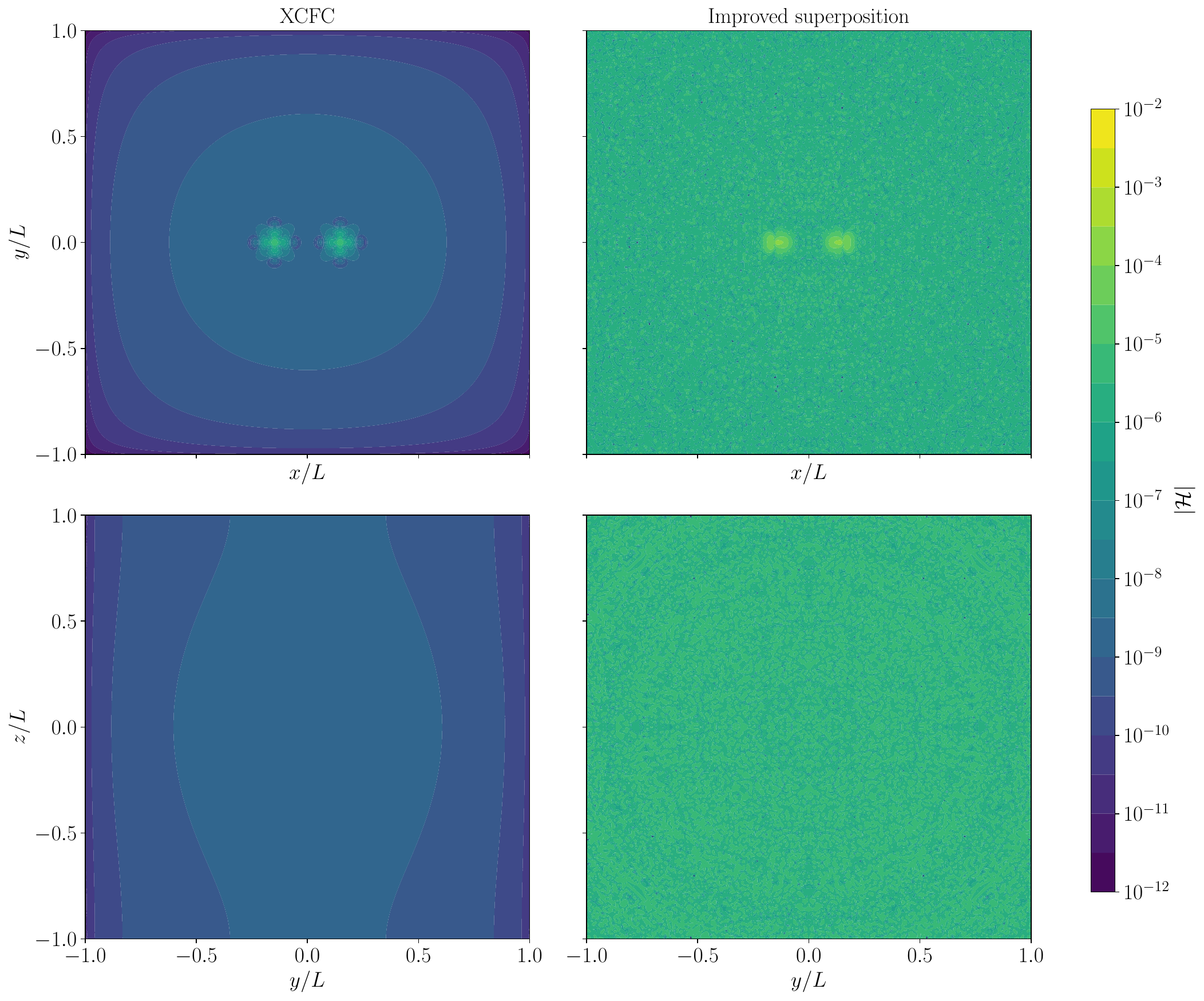} 
    \caption{\textit{Left panels:} Hamiltonian constraint violation on the $xy$-plane (top) and $yz$-plane (bottom) for an equal-mass boson star binary built with our code using XCFC. \textit{Right panels:} Same as the left panel but using the improved superposition approach. In all plots the resolution is $N_p^3=250^3$ and central scalar field value is  $\phi_0=3.0\times10^{-2}$. The Cartesian axes have been  normalized with respect to $L=10R_{99}$.
}
    \label{fig:BBS_ham_plane}
\end{figure*}

Boson stars are horizonless, localized, self-gravitating solitonic configurations, characterized by a discrete eigenfrequency $\omega/\mu$. These objects arise as solutions of the Einstein--(complex, massive) Klein--Gordon system in the case of scalar boson stars~\cite{Schunck:2003kk}, and of the Einstein--(complex) Proca system in the case of vector boson stars (also known as Proca stars)~\cite{BRITO2016291}. The interested reader is addressed to~\cite{Liebling_2012,bezares2025exotic} for recent reviews on such exotic compact objects.

In this work, we focus exclusively on scalar boson stars, which are solutions of the Einstein--(complex, massive) Klein--Gordon system,
\begin{eqnarray}
    G_{\mu\nu} &=& 8\pi T_{\mu\nu},
\\
    g^{\mu\nu} \nabla_{\mu} \nabla_{\nu} \Phi &=& \Phi \frac{dV}{d|\Phi|^2},
\end{eqnarray}
where the stress-energy tensor $T_{\mu\nu}$ is given in Eq.~(\ref{eq:Tmunu_field}). Boson stars can exist in different configurations depending on the choice of the scalar potential $V\left(|\Phi|^2\right)$. Here, we restrict our analysis to \textit{mini}-boson stars, for which the potential corresponds to that of a free massive scalar field,
\begin{equation}
    V\left(|\Phi|^2\right) = \mu^2 |\Phi|^2.
\end{equation}
We note that the particle mass $\mu$ plays the role of a fundamental scale of the system, and each physical quantity can be appropriately rescaled by $\mu$. Here, we set $\mu = 1$.

The scalar field ansatz for boson stars in spherical coordinates is
\begin{equation}
    \Phi(t,r,\theta,\varphi) = \phi(r,\theta)\, e^{i(\omega t - m \varphi)},
\end{equation}
where $\phi$ denotes the scalar field amplitude, $\omega$ is the eigenfrequency, and $m$ is an integer azimuthal quantum number. At large distances, the scalar field decays exponentially, ensuring that spacetime is asymptotically flat. We restrict our analysis to spherically symmetric solutions, corresponding to non-rotating boson stars ($m = 0$), with a non-vanishing central value of the scalar field, denoted by $\phi_0$. Since boson stars do not have a surface discontinuity of the energy density (contrary to fermionic stars), their compactness is usually defined as the ratio of the radius $R_{99}$ that contains $99\%$ of the mass $M_{99}$, $C=2M_{99}/R_{99}$. Further details on the construction of these isolated configurations can be found in~\cite{Liebling_2012,bezares2025exotic}.

The top left panel of Fig.~\ref{fig:BBS_all} shows the mass--frequency diagram for our boson star model (solid black curve). Overlaid in this curve, the red dots indicate the specific configurations considered in this study, which are generated using the spectral-method code developed in~\cite{Lazarte:2023a,Lazarte_2024}.
Those lie on the stable branch, corresponding to masses below the maximum value ($M\mu\sim0.63$). We build constraint-satisfying initial data for equal-mass binary boson stars and compare the solutions obtained by our formalism with alternative approaches.

\subsection{The no initial boost static case}
\label{subsec:BBS_noboost}

Initial data for binary boson stars are often constructed using a simple superposition of two isolated star solutions. Let us consider two boson stars with centers located at $x_A^i$ and $x_B^i$, respectively. The spacetime of each isolated star is described by the spatial metrics $\gamma_{ij}^X$, where $X$ denotes the star. The spatial metric of the binary system is then constructed from the individual metrics as
\begin{equation}
\gamma_{ij} = \gamma_{ij}^A + \gamma_{ij}^B - f_{ij}\,.
\end{equation}

This leading-order approximation becomes exact in the limit in which the two stars are infinitely separated. However, as shown in~\cite{Helfer_2022}, the absence of horizons in boson star binaries can lead to significant artifacts when the spacetime is constructed via simple superposition. In particular, the scalar field of each star exhibits spurious pulsations during the subsequent time evolution. 

A modification of this approach, applicable to equal-mass binaries, was first suggested in~\cite{PhysRevD.99.044046}. In this construction, the simple superposition is replaced by what we refer to as improved superposition, defined as
\begin{equation}
\gamma_{ij} = \gamma_{ij}^A + \gamma_{ij}^B - \gamma_{ij}^B\left(x_A^i\right) = \gamma_{ij}^A + \gamma_{ij}^B - \gamma_{ij}^A\left(x_B^i\right),
\label{improved}
\end{equation}
where the two expressions are equivalent for equal-mass configurations.

This prescription ensures that, at the location of each star’s center, the corresponding equilibrium spatial metric and volume element are exactly recovered (overcoming the spurious pulsating effects observed in~\cite{Helfer_2022}), with the drawback that the asymptotic metric no longer approaches the flat metric $f_{ij}$. We also note that  prescription (\ref{improved}) does not work for asymmetric configurations such as binaries with unequal masses or spins, but a similar prescription was proposed for the unequal case in~\cite{evstafyeva2023unequal}.

We use our code in a fully three-dimensional setting in Cartesian coordinate with a non-staggered grid, with both a convergence and relaxation threshold of $10^{-10}$. This corresponds to a good compromise between minimizing numerical spurious effects and maintaining a reasonable computational cost. In order to further improve the latter, we employ an MPI-based parallelization scheme with $5$ processors for each spatial direction. As in the toroidal toy model, Dirichlet boundary conditions are imposed on both the inner (parallel) domain boundaries and the outer boundaries of the computational grid. As initial guesses for the metric functions, we consider both simple and improved superposition approaches. However, in the following analysis we focus exclusively on the former. 

The computational domain extends over a region $20$ times the radius $R_{99}$ of each star, such that all variables are defined in the interval $[-L, L]$ with $L = 10 \,R_{99}$. We fix the stars' separation to $D=40$ (locating the stars along the $x$-axis at position $x^i=\pm20$). No initial boost is added to either star. To compare  results for different resolutions, we choose $N_p^3=\left(200^3, 250^3\right)$. The number of grid points plays a crucial role in the convergence of the code: additional tests (not shown here) indicate that lower resolution fails to achieve convergence for the most compact configurations.

The top right panel of Fig.~\ref{fig:BBS_all} displays the $L_2$-norm of the Hamiltonian constraint violation. 
The (constraint-satisfying) results obtained with XCFC are shown both, in the solid black line ($N_p^3 = 200^3$), and in the dashed red line ($N_p^3 = 250^3$, rescaled for a convergence order of $3.7$). Those are compared with the (constraint-violating) results obtained from simple superposition (dashed green line) and improved superposition (dashed blue line), both computed at a resolution of $N_p^3 = 250^3$. As theoretically expected, the black and red curves overlap with nearly fourth-order convergence, except for small deviations in the less compact configurations. This behavior arises because, although fourth-order convergence of the Hamiltonian constraint violation is recovered around the stellar positions, it is lost once the violation reaches the floor value. The $L_2$-norm of the Hamiltonian constraint violation spans a range in between $10^{-6}$ (for less compact configurations) and slightly above $10^{-4}$ (for more compact ones). It is particularly worth mentioning that the black and red curves always lie well below the superposition curves, indicating that our constraint-satisfying initial data outperform those obtained through superposition approaches, even when the latter are computed at higher resolution.

The bottom-left panel of Fig.~\ref{fig:BBS_all} shows the computational time required for convergence (in seconds) as a function of the stars' frequency, and therefore their compactness. We find that increasing compactness results in longer convergence times. Specifically, the time required to reach convergence grows from a few tens of minutes to nearly $18$ hours for the lower resolution, and from more than an hour up to almost $2$ days for the higher resolution (see also Table~\ref{tab:time}). Due to this substantial runtime increase, we restrict our analysis to low-compactness, stable-branch stars and adopt a moderate threshold.

Correspondingly, the bottom-right panel of Fig.~\ref{fig:BBS_all} shows the number of iterations required to achieve convergence when solving the metric variables with a resolution of $N_p^3 = 200^3$. The results for the less compact binary ($\phi_0 = 5.0 \times 10^{-3}$) are shown in black (lapse function) and in red (conformal factor), while the respective results for the most compact binary ($\phi_0 = 4.5 \times 10^{-2}$) are shown in green and blue lines.  As  compactness increases, more iterations are required to reach convergence (from less than $10$ to almost $900$). We note that the resolution does not significantly affect the number of iterations, which remains approximately constant; however, the computational time increases due to the larger number of points that the code must handle.
Faster configurations can be achieved depending on the chosen resolution. The main drawback, as already mentioned, is that an insufficient resolution does not allow the most compact configurations to be properly resolved.
Since the results obtained at $N_p^3 = 200^3$ resolution are superior to those produced by superposition schemes at even higher resolution, we conclude that our code can be operated in a regime that provides an optimal compromise between computational cost and the quality of the results.

Figure~\ref{fig:BBS_ham_plane} shows the Hamiltonian constraint violation obtained for an equal-mass boson star binary with $\phi_0=3.0\times10^{-2}$ and a separation of $D=40$, both at the $xy$-plane (top panels) and at the $yz$-plane (bottom panels). The right panels display the results for constraint-satisfying initial data for a grid resolution $N_p^3=250$ while the left panels those for the improved superposition method. We have normalized the axes with respect to $L=10\,R_{99}$. The results obtained through superposition exhibit a floor error of the order of $10^{-7}$ (with some noise), while those obtained with XCFC reach values as low as $10^{-12}$. A gain of the same order of magnitude could also be achieved in the vicinity of the stars’ locations, indicating that our XCFC approach is capable of producing significantly more accurate initial data for binary boson star systems.

\begin{figure}
    \centering
    \includegraphics[width=0.45\textwidth]{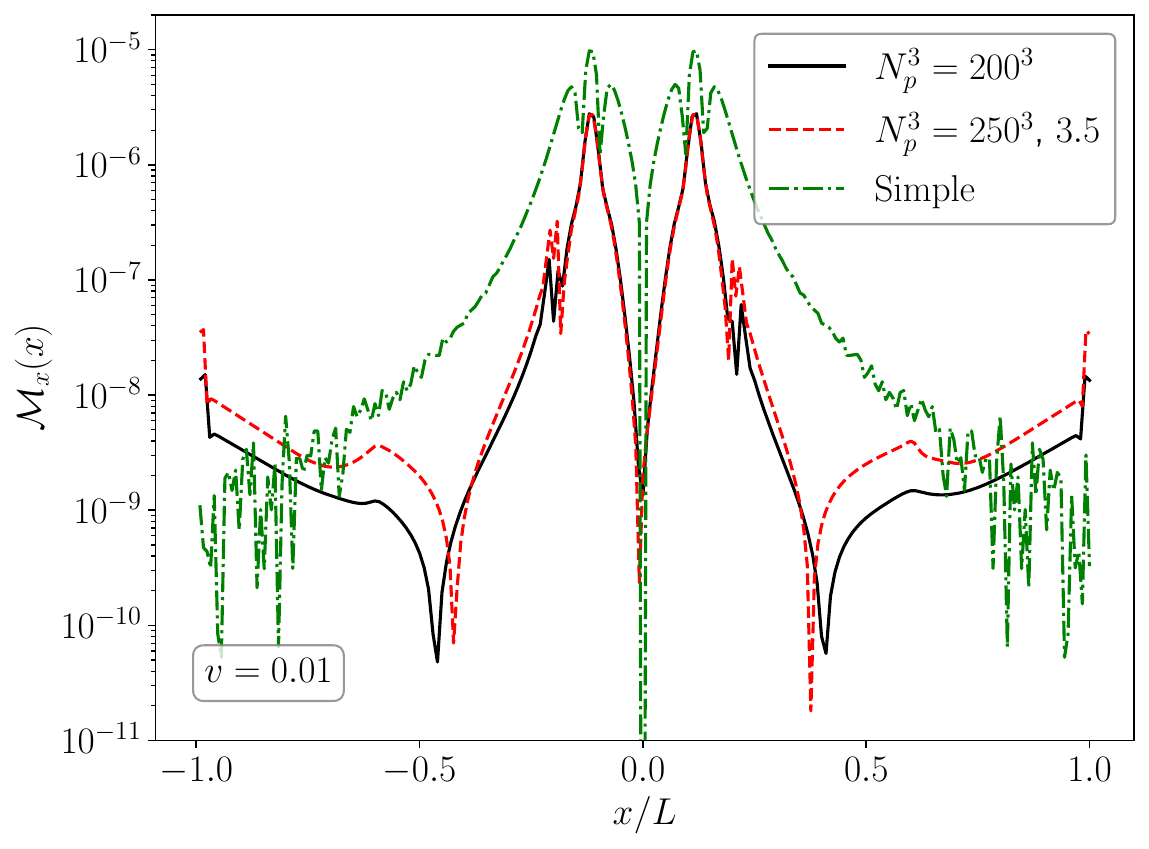} 
    \includegraphics[width=0.45\textwidth]{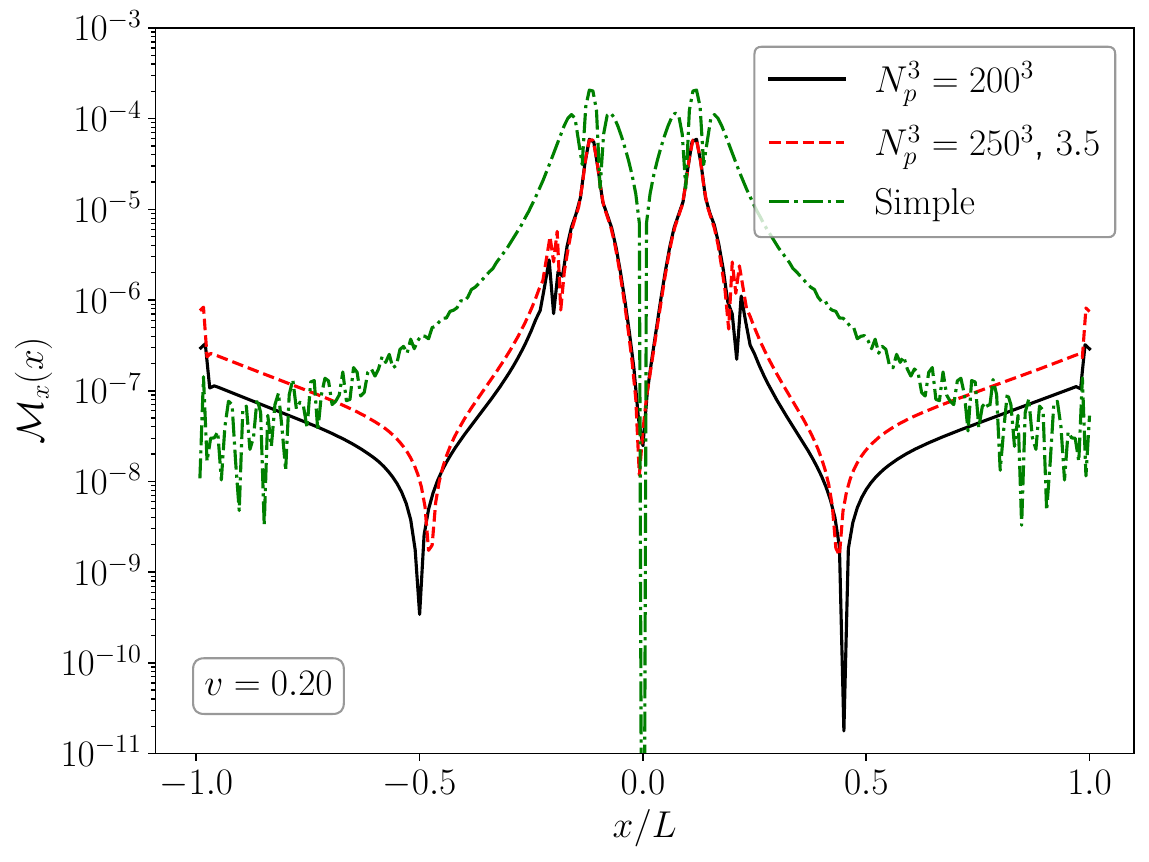}
    \caption{Convergence test for the violation of the $x$-component of the momentum constraint ($\mathcal{M}_x$) along the $x$-axis for an equal-mass binary boson star ($\phi_0=2.0\times 10^{-2}$ and initial boosts $v=0.01$ (top) and $v=0.20$ (bottom)), for grid resolutions, $N_p^3 = (200^3, 250^3)$ (solid black and dashed red lines, respectively). The latter is rescaled according to a convergence order of $3.5$. Dashed green curves indicate the violation of $\mathcal{M}_x$ obtained through simple superposition (for $N_p^3 =250^3$). In both plots, the $x$-axis has been rescaled by $L = 10\,R_{99}$. 
}
    \label{fig:boosted_BBS_Momx}
\end{figure}

\begin{figure*}[t] 
    \centering
    \includegraphics[width=1.025\textwidth]{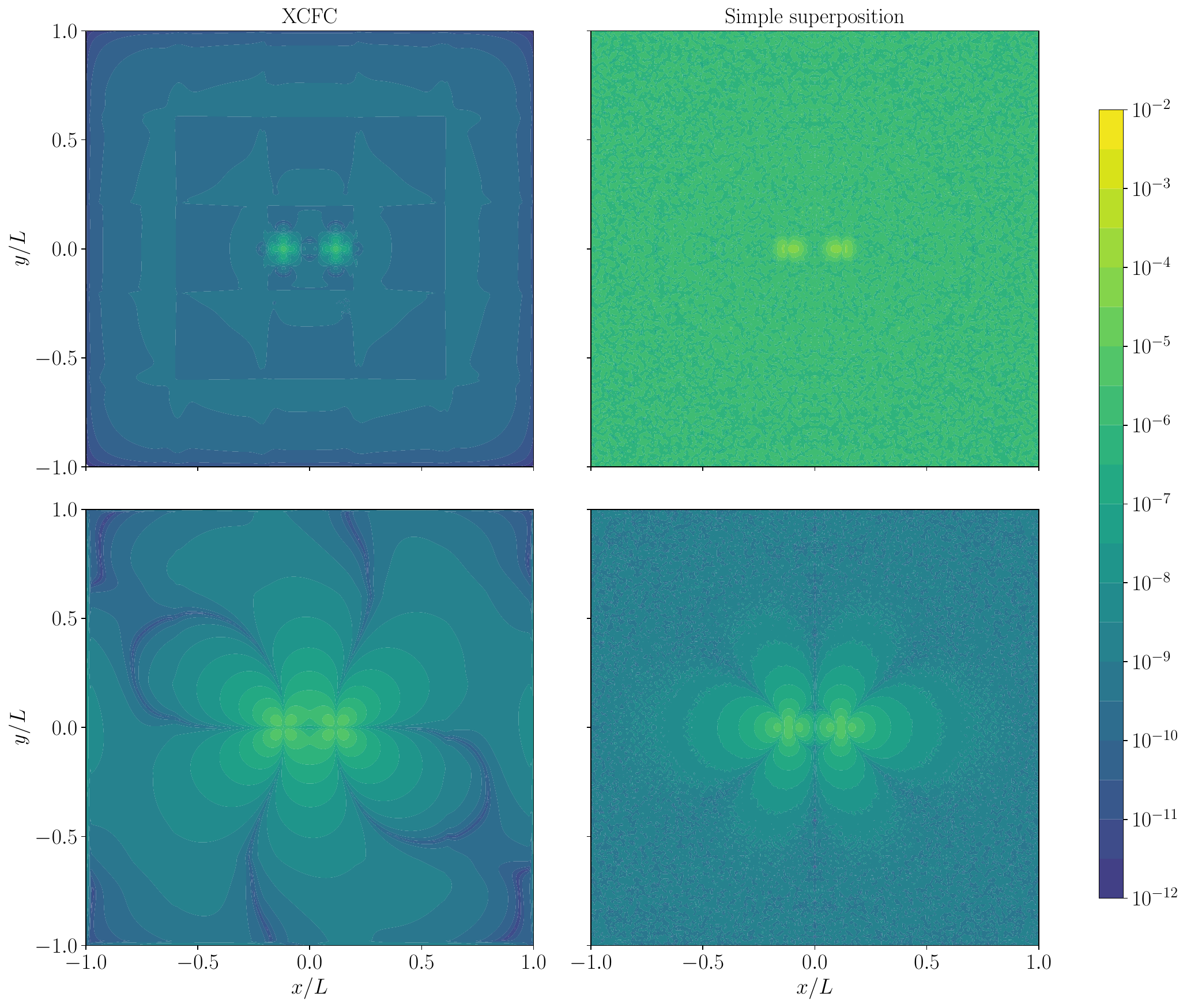} 
    \caption{\textit{Top panels:} Hamiltonian constraint violation on the $xy$-plane for an equal-mass boson star binary with $\phi_0=2.0\times10^{-2}$, initial boost $v=0.01$, and grid resolution $N_p^3=250^3$. The left plot sows the result for the XCFC approach and the right one for simple superposition. \textit{Bottom panels:} same as above but for the $x$-component of the momentum constraint. The Cartesian axes have been normalized with respect to $L=10\,R_{99}$.
}
    \label{fig:BBS_ham_mom_plane_0.01}
\end{figure*}

\begin{figure*}[t] 
    \centering
    \includegraphics[width=1.025\textwidth]{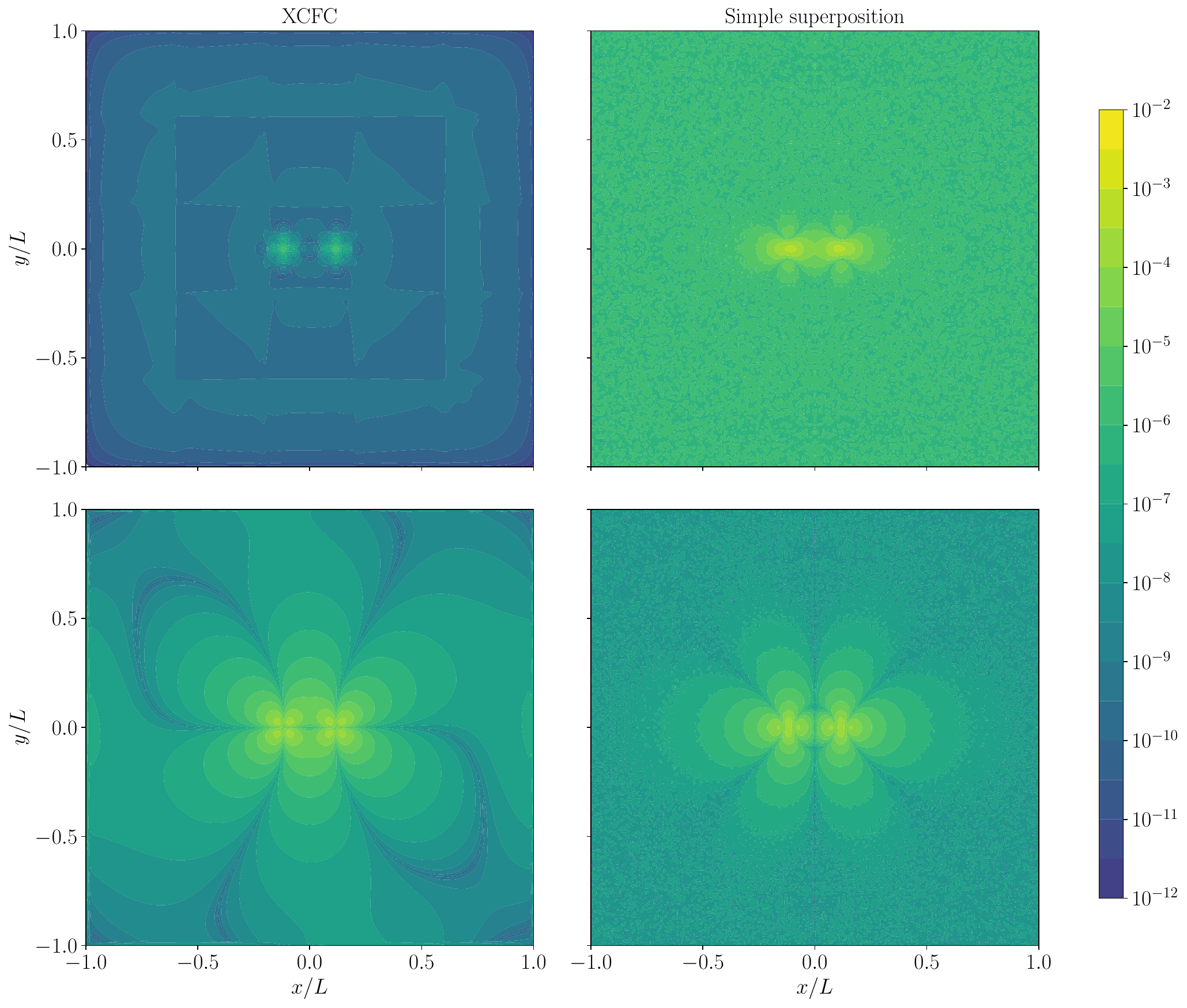} 
    \caption{Same as Fig.~\ref{fig:BBS_ham_mom_plane_0.01} but for and initial boost of $v=0.20$. 
}
    \label{fig:BBS_ham_mom_plane_0.20}
\end{figure*}

\subsection{The initial boost case}
\label{subsec:BBS_boost}

We turn now to consider the same binary boson star configurations discussed in the previous section but including initial linear momentum. This is achieved by boosting each star with a spatial velocity $\vec{v}$. 
Such a configuration leads to a non-trivial computation of the momentum constraint. In our analysis we focus on stars boosted towards each other. As before, we fix the stellar separation to $D = 40$, placing the stars along the $x$-axis at positions $x_A^i = 20$ and $x_B^i = -20$. Consequently, the assigned spatial velocities are $\vec{v}_A = (-v,0,0)$ and $\vec{v}_B = (v,0,0)$, respectively. To construct the binary configurations we start from isolated star solutions and apply a Lorentz transformation corresponding to the prescribed boost. As an initial guess for our solver, we simply superpose these isolated Lorentz-transformed solutions.

As before, we adopt a fully three-dimensional setting in Cartesian coordinates based on a non-staggered grid and set both relaxation and convergence thresholds to $10^{-10}$. This choice represents a reasonable compromise between minimizing numerical artifacts and maintaining an acceptable computational cost. However, this configuration proved insufficient to achieve convergence. We therefore split the iterative procedure into two stages, introducing an additional outer-iteration threshold fixed at $10^{-10}$. In the first stage, the code solves for the lapse function and the conformal factor, iterating until the residuals of the corresponding elliptic equations fall below the prescribed relaxation threshold and global convergence for these variables is achieved. Once this condition is satisfied, the iteration for the shift vector is activated and carried out until convergence (second stage). To ensure the overall consistency of the solution, the code then alternates between these two stages, repeating the full cycle until global convergence is attained, namely when the solution stabilizes within the prescribed outer-iteration threshold. The computational efficiency is further improved through an MPI-based parallelization scheme, using $5$ processors along each spatial direction. As before, we impose Dirichlet boundary conditions at both the inner (parallel) domain boundaries and the outer boundaries of the grid.

The computational domain extends over a region of size $20$ times the radius $R_{99}$ of each star, such that all variables are defined in the interval $[-L,L]$, with $L = 10\,R_{99}$. In order to compare the results at different resolutions, we consider, as before, $N_p^3 = \left(200^3,\,250^3\right)$. Our particular system is an equal-mass binary boson star system with $\phi_0 = 2.0 \times 10^{-2}$. We select several values of the initial boost, namely $v = [0.01, 0.10, 0.15, 0.20]$, for both resolutions. This choice implies a significant computational cost (see Table~\ref{tab:time}), ranging from approximately $3$ hours for the lowest boost up to nearly $21$ hours for the highest boost at low resolution ($N_p^3 = 200^3$). For the higher resolution ($N_p^3 = 250^3$), the computational time increases substantially, spanning from about $9$ hours to nearly $2.5$ days. 

In the following, we focus on the configurations with the lowest and highest boost, namely $v = 0.01$ and $v = 0.20$. 
Fig.~\ref{fig:boosted_BBS_Momx} shows a convergence test for the $x$-component of the momentum constraint violation ($\mathcal{M}_x$) along the $x$-axis (rescaled by $L = 10\,R_{99}$). The top plot corresponds to $v=0.01$ and the bottom one to $v=0.20$, and both show the two resolutions ($N_p^3 = 200^3$ (solid black line), $N_p^3 = 250^3$ (dashed red line)). The higher-resolution result has been rescaled according to a convergence order of $3.5$.
In addition, the dashed green line shows the  violation of $\mathcal{M}_x$ that is obtained when using simple superposition at the higher resolution. Convergence is well recovered in the central regions close to the positions of the stars, while deviations appear near the boundaries. This reveals the impact of boundary effects discussed in Section~\ref{subsec:gaussian_scalar_field_profile} and further analyzed in Appendix~\ref{app:1}. Once again, the comparison with the simple superposition approach shows that our method outperforms it, achieving lower constraint violations even at lower resolution.

Figures~\ref{fig:BBS_ham_mom_plane_0.01} and~\ref{fig:BBS_ham_mom_plane_0.20} show the violations of the Hamiltonian constraint (top panels) and of the $x$-component of the momentum constraint (bottom panels) for $v=0.01$ and $v=0.20$, respectively. The results are shown on the equatorial plane at a grid resolution of $N_p^3 = 250^3$. The XCFC results are displayed in the left plots while those obtained via simple superposition are shown in the right ones. The Hamiltonian constraint violation, characterised in the top-left plot by a grid-like structure arising from the relatively high threshold values selected, reaches a floor value of approximately $10^{-12}$ with our XCFC data and does not exceed a maximum value of $10^{-6}$, attained at the stellar positions. In contrast, the superposition approach reaches maximum Hamiltonian violation values of about $10^{-3}$ at the stellar positions and a floor of approximately $10^{-7}$. On the other hand, the momentum constraint violation accomplished by the XCFC solver shows a less significant improvement to simple superposition, reaching values of order $10^{-3}$ at the stars' locations and a lower floor value ($10^{-9}$). It is worth highlighting, however, the absence of noisy oscillations in the XCFC solution as compared to the superposition approach (due to the approximation and errors introduced in the latter).

\section{Conclusions}

In this work, we have discussed the problem of constructing constraint-satisfying initial data for numerical relativity simulations, with particular emphasis on scalar-field configurations and binary boson star systems, ensuring the physical consistency of the resulting spacetimes with Einstein's equations of General Relativity. Our study is motivated by the current state of the art, in which constraint-violating initial data are still widely used even though several codes able to produce constraint-satisfying initial data exist and a number of studies employing such data have recently been conducted~\cite{Aurrekoetxea_2023,PhysRevD.107.124018,
PhysRevD.108.124015,PhysRevD.109.044058,aurrekoetxea2025grtresnaopensourcecodesolve}. Our approach joins those efforts, providing a robust tool for generating constraint-satisfying initial data. In particular, we have explored the use of the XCFC formalism~\cite{cordero-carrion2008}. This formulation has been shown to guarantee existence and uniqueness of the solution, thus providing a robust foundation for the construction of physically consistent initial data.

We have developed a new numerical code, \textsc{Incipit}, and implemented our initial-data solver within the XCFC framework, validating its performance through a series of tests. Specifically, we have verified the convergence and stability of the code starting from simplified toy models, such as complex scalar-field configurations with Gaussian and toroidal profiles. Our results have shown that the hierarchical elliptic structure of XCFC leads to well-behaved solutions and avoids the non-uniqueness issues reported in other formulations~\cite{PhysRevLett.95.091101,PhysRevD.75.044009, Rinne_2008}.
We have then applied our solver to build constraint-satisfying initial data for binary boson star systems, considering both static and boosted configurations. The comparison with  approaches based on the superposition of isolated star solutions has revealed that our method outperforms those strategies. Our approach can reduce spurious effects in the early stages of the evolution, thereby improving the physical consistency of the simulations.

More generally, the approach presented here can be regarded as a viable alternative to existing methods for the construction of initial data in General Relativity, especially in the context of boson star binaries and, in a broader sense, compact objects involving nontrivial matter configurations. In this work, we have demonstrated the versatility of our initial-data solver, which can be adapted to a variety of configurations, including rotating objects such as scalar-field configurations with toroidal topology. A current limitation of our approach is the computational cost required to obtain convergent solutions, even if MPI parallelization is used. This is mainly due to the high resolution demanded by the use of a uniform grid, which is essential for accurately solving the elliptic system. Future developments in this line of research will focus on a more extensive validation of the method, detailed comparisons with other formalisms, and extensions beyond the conformally flat approximation. In addition, improvements regarding the numerical implementation, such as moving beyond a uniform grid, will be explored in order to reduce the computational cost. 

\begin{acknowledgments}
GP acknowledges support from the Spanish Agencia Estatal de Investigaci\'on through grant PRE2022-104185
funded by MICIU/AEI/10.13039/501100011033 and by FSE+.
This work is supported by the Spanish Agencia Estatal de Investigaci\'on (grants PID2024-159689NB-C21 and PID2022-136828NB-C43 funded by MICIU/AEI/10.13039/501100011033 and by FEDER / EU),  
by the Generalitat Valenciana through the Prometeo program for excellent research groups (grant CIPROM/2022/49) and Santiago Grisolía Grant (CIGRIS/2022/164), and by the European Horizon Europe staff exchange (SE) programme 
HORIZON-MSCA-2021-SE-01 (Grant No.-NewFunFiCO-101086251) and 
HORIZON-MSCA-2024-SE-01 (Grant No.-Gravity-101236384). NSG acknowledges support from the Spanish Ministry of Science, Innovation, and Universities via the Ram\'on y Cajal programme (grant RYC2022-037424-I), funded by MICIU/AEI/10.13039/501100011033 and by ESF+. The authors acknowledge computer resources provided by the Red Espa\~nola de Supercomputaci\'on (Tirant, MareNostrum5, Altamira, and Storage5) and the technical support from the IT departments of the Universitat de Val\`encia and the Barcelona Supercomputing Center (Projects No.~RES-FI-2024-2-0012 and No.~RES-FI-2024-3-0007) and by Instituto de Física de Cantabria  (IFCA) (Altamira) through Project No.~FI-2025-1-0011). Computational resources were also provided via the Portuguese Foundation for Science and Technology (FCT, \url{https://ror.org/00snfqn58}) through project 2025.09498.CPCA.A3. We thank FDG for useful suggestions and valuable discussion.
\end{acknowledgments}

\appendix
\section{The role of boundary conditions and threshold on code convergence}
\label{app:1}

\begin{figure}[t] 
    \centering
    \includegraphics[width=0.45\textwidth]{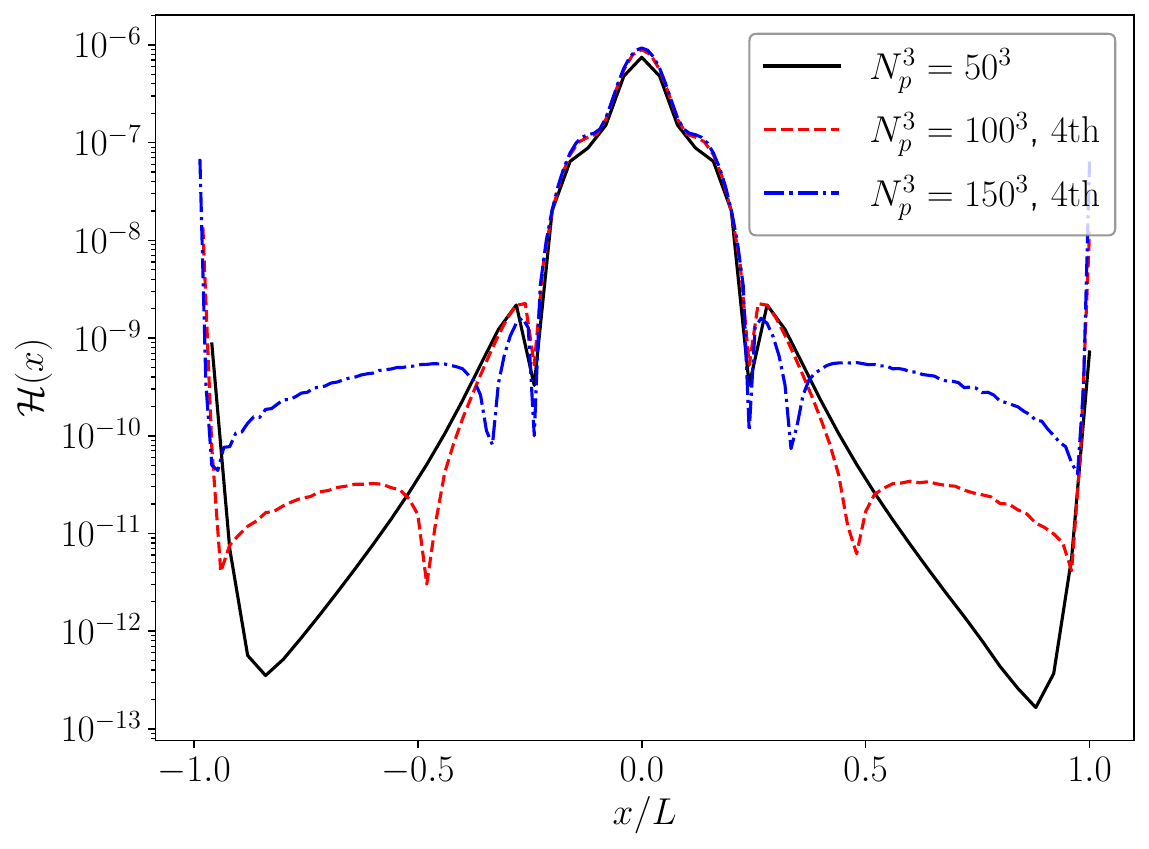}
    \includegraphics[width=0.45\textwidth]{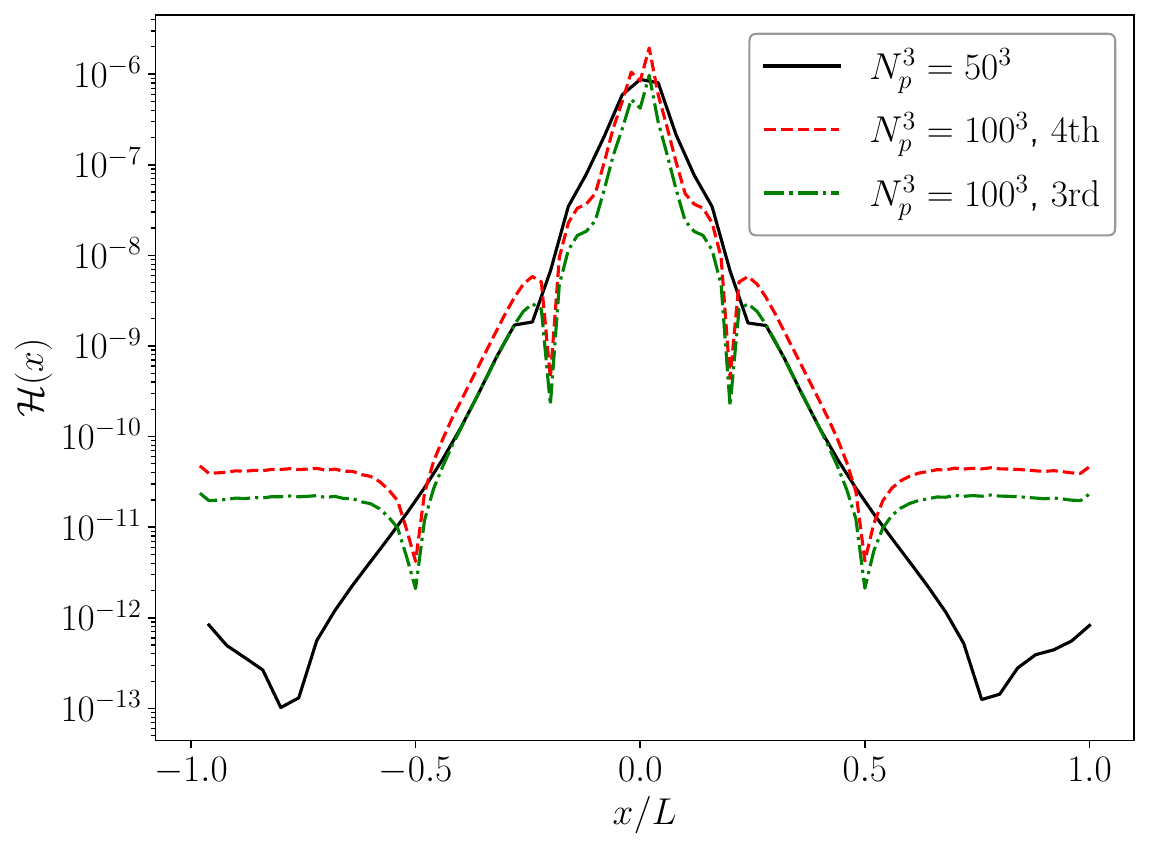}
    \caption{Convergence for the Hamiltonian constraint violation with a Gaussian-like scalar field profile, along the $x$-axis which has been conveniently normalized with respect to $L=10\sigma$. \textit{Top panel:} Using Dirichlet boundary condition and a threshold equals to $10^{-14}$. Different resolutions have been implemented, corresponding to $N_p^3=(50^3,100^3,150^3)$, shown as solid black, dashed red, and dashed green lines, respectively. Each line has been rescaled according to fourth-order convergence. \textit{Bottom panel:} Using Robin boundary condition and a threshold equals to $10^{-14}$. Just two resolutions have been displayed $N_p^3=(50^3,100^3)$ shown as black and dashed red lines rescaled according fourth-order convergence. The dashed green curve represents the rescaled constraint violation assuming a convergence order of $3.5$}
    \label{fig:BC_analisys}
\end{figure}

In this Appendix we examine the effects of the choice of boundary conditions and of the convergence and relaxation thresholds on the convergence test performed for a Gaussian-like scalar field profile. The bottom panel of Fig.~\ref{fig:gaussian_hamiltonian_constraint_convergence} shows fourth-order convergence obtained using Robin boundary conditions and a threshold of $10^{-15}$. While an inappropriate choice of boundary conditions may introduce boundary-related artifacts, a permissive convergence threshold leads to insufficiently converged solutions, introducing a tolerance-induced error. This error is not related to the numerical solver itself nor to the finite-difference discretizations implemented in the code.

The top panel of Fig.~\ref{fig:BC_analisys} shows the convergence test performed for the same model presented in Sec.~\ref{subsec:gaussian_scalar_field_profile}, but with Dirichlet boundary conditions and a $10^{-14}$ convergence threshold. The Hamiltonian constraint violation profiles for each resolution, $N_p^3=\left(50^3,100^3,150^3\right)$, represented by solid black, dashed red and dotted-dashed blue lines, respectively, have been rescaled according to fourth-order convergence. Convergence is observed only in the central region. Near the boundaries spurious effects affect the convergence, which is no longer maintained. In general, poor choice of boundary conditions results in reflections of the constraint violations at the boundaries (see also \cite{Alcubierre_2008}), leading, in our case, to a divergence of the violation near the boundaries, reaching the same order of magnitude as the central error. 

Keeping the same convergence threshold, we switch to Robin boundary conditions, which, in asymptotically flat spacetimes with a power-law fall-off, are commonly adopted to mitigate boundary effects while preserving the well-posedness of the problem~\cite{Thornburg_1987,  Cook_2000, LAU20071126}. For each face of the finite-volume computational domain, we assume that the variable $u$ decays as $u \sim M/r^n$,
where $r$ is the radial coordinate, defined in Cartesian coordinates as $r = \sqrt{x^2 + y^2 + z^2}$. The decay order is fixed to $n = 1$ for scalar variables and $n = 2$ for vector variables.
The bottom panel of Fig.~\ref{fig:BC_analisys} shows the convergence test for this model. The curves corresponding to the resolutions $N_p^3 = (50^3, 100^3)$, shown in solid black and dashed red lines, respectively, have been rescaled assuming fourth-order convergence. The dashed green curve represents the rescaled constraint violation assuming a convergence order of $3.5$. 
The figure shows that convergence is not maintained near the boundaries. In those regions the floor error dominates, where the violation is smaller, degrading the convergence. However, the Robin boundary conditions appear to be consistent with the problem under investigation, since the numerical domain is sufficiently large for the variable to lie in a radially-decaying region.

\bibliography{Biblio}

\end{document}